\documentclass[utf8]{FrontiersinVancouver} 

\usepackage{url,hyperref,lineno,microtype}
\usepackage[onehalfspacing]{setspace}

\usepackage{booktabs}

\usepackage{tikz}
\usetikzlibrary{positioning,calc,arrows.meta}
\usepackage{pgfplots}
\pgfplotsset{compat=1.18}
\usepackage{pgf-pie}
\usepackage{graphicx}
\usepackage{grffile}   
\usepackage{array}     

\def\keyFont{\fontsize{8}{11}\helveticabold }
\def\firstAuthorLast{Rehmat {et~al.}} 
\def\Authors{Sara Rehmat\,$^{1}$, Hafeez Ur Rehman\,$^{1,2,*}$,Byeong-Gwon Kang\,$^{3,*}$, Sarra Ayouni\,$^{4}$, Yunyoung Nam\,$^{3}$}
\def\Address{$^{1}$Department of Computer Science, National University of Computer and Emerging Sciences, Islamabad, Pakistan\\
$^{2}$College of Computing and Data Science, Oryx University with Liverpool John Moores University, Doha, Qatar\\
$^{3}$Department of ICT Convergence, Soonchunhyang University, South Korea\\
$^{4}$Department of Information Systems, College of Computer and Information Sciences, Princess Nourah bint Abdulrahman University, P.O. Box 84428, Riyadh 11671, Saudi Arabia.
}
\def\corrAuthor{Hafeez Ur Rehman, Byeong-Gwon Kang}

\def\corrEmail{h.urrehman@ljmu.ac.uk, hafeez.urrehman@nu.edu.pk, bgkang@sch.ac.kr}

\begin{document}
\onecolumn
\firstpage{1}

\title[Variance Penalized GAN for HE to IHC Transformation]{Transforming Hematoxylin-Eosin Images into Immunohistochemistry Images: A Variance-Penalized GAN for Precision Oncology} 

\author[\firstAuthorLast ]{\Authors} 
\address{} 
\correspondance{} 

\extraAuth{}
\maketitle
\begin{abstract}
The overexpression of the human epidermal growth factor receptor 2 (HER2) in breast cells is a key driver of HER2-positive breast cancer, a highly aggressive subtype requiring precise diagnosis and targeted therapy. Immunohistochemistry (IHC) is the standard technique for HER2 assessment but is costly, labor-intensive, and highly dependent on antibody selection. In contrast, hematoxylin and eosin (H\&E) staining, a routine histopathological procedure, offers broader accessibility but lacks HER2 specificity. This study proposes an advanced deep learning-based image translation framework to generate high-fidelity IHC images from H\&E-stained tissue samples, enabling cost-effective and scalable HER2 assessment. By modifying the loss function of pyramid pix2pix, we mitigate mode collapse, a fundamental limitation in generative adversarial networks (GANs), and introduce a novel variance-based penalty that enforces structural diversity in generated images. Our model particularly excels in translating HER2-positive (IHC 3+) images, which have remained challenging for existing methods. Quantitative evaluations on the overall BCI dataset reveal that our approach outperforms baseline models, achieving a peak signal-to-noise ratio (PSNR) of 22.16, a structural similarity index (SSIM) of 0.47, and a Fréchet Inception Distance (FID) of 346.37. In comparison, the pyramid pix2pix baseline attained PSNR 21.15, SSIM 0.43, and FID 516.75, while the standard pix2pix model yielded PSNR 20.74, SSIM 0.44, and FID 472.6. These results affirm the superior fidelity and realism of our generated IHC images. Beyond medical imaging, our model exhibits superior performance in general image-to-image translation tasks, showcasing its potential across multiple domains. This work marks a significant step toward AI-driven precision oncology, offering a reliable and efficient alternative to traditional HER2 diagnostics.  

\tiny
 \keyFont{ \section{Keywords:} breast cancer, HER2 cancer, image translation, GANs, mode collapse} 
\end{abstract}

\section{Introduction}

Breast cancer remains the most prevalent form of cancer worldwide \cite{lei2021global}. Among its various subtypes, approximately 20\% of the cases are classified as HER2-positive breast cancer, characterized by overexpression of the human epidermal growth factor receptor 2 (HER2) \cite{Siegel2022}. This subtype is particularly aggressive and requires targeted therapies, such as Herceptin, to inhibit HER2 signaling. However, accurate HER2 classification is critical, as misdiagnosis can lead to inappropriate treatment strategies and higher recurrence rates \cite{slamon1987human}.

\begin{figure}[htbp]
\centering

\begin{minipage}{0.45\linewidth}
\centering
\includegraphics[width=\linewidth]{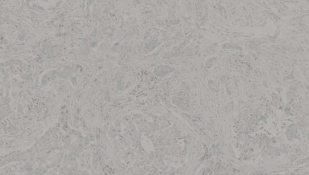}
\small (a) IHC 0
\label{fig:ihc0}
\end{minipage}
\hfill
\begin{minipage}{0.45\linewidth}
\centering
\includegraphics[width=\linewidth]{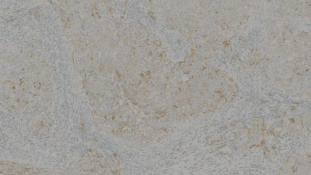}
\small (b) IHC 1+
\label{fig:ihc1}
\end{minipage}

\vspace{0.5cm}

\begin{minipage}{0.45\linewidth}
\centering
\includegraphics[width=\linewidth]{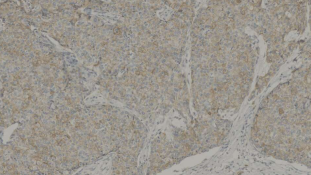}
\small (c) IHC 2+
\label{fig:ihc2}
\end{minipage}
\hfill
\begin{minipage}{0.45\linewidth}
\centering
\includegraphics[width=\linewidth]{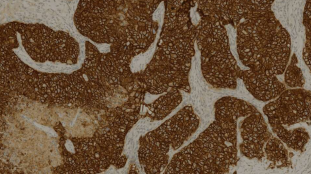}
\small (d) IHC 3+
\label{fig:ihc3}
\end{minipage}

\caption{Different types of IHC images based on levels of HER2 expression in the BCI dataset \cite{liu2022bci}. The higher the HER2 expression, the greater the corresponding label.}
\label{fig:IHC_Types}
\end{figure}

The gold standard for HER2 detection is fluorescence in situ hybridization (FISH), a molecular technique that assesses gene amplification \cite{prati2005histopathologic}. However, FISH is costly, time consuming and not widely available in all clinical laboratories \cite{ross2004targeted}. Immunohistochemistry (IHC) provides a more accessible alternative, relying on antibody-antigen interactions to detect HER2 expression \cite{thomson2001her,van2001assessment}. IHC results are categorized into four levels: 0, 1+, 2+, and 3+, as illustrated in Figure \ref{fig:IHC_Types}. Although IHC 0 and 1+ are HER2-negative, IHC 3+ is strongly HER2-positive and indicates eligibility for HER2-targeted therapies. However, IHC 2+ remains ambiguous, necessitating additional FISH testing \cite{Siegel2022}. Despite its utility, IHC still requires specialized reagents, expensive equipment, and experienced pathologists for interpretation, introducing subjectivity in diagnosis \cite{robb2015call,lin2014standardization}.

\begin{figure}[htbp]
\centering
\begin{minipage}{0.32\linewidth}
\centering
\includegraphics[width=\linewidth]{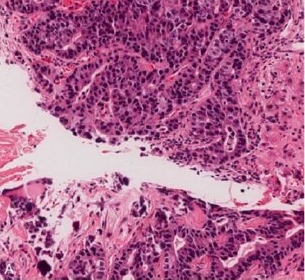}
\small (a) H\&E
\end{minipage}\hfill
\begin{minipage}{0.32\linewidth}
\centering
\includegraphics[width=\linewidth]{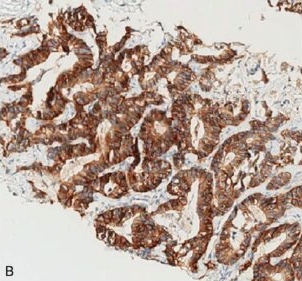}
\small (b) IHC
\end{minipage}\hfill
\begin{minipage}{0.32\linewidth}
\centering
\includegraphics[width=\linewidth]{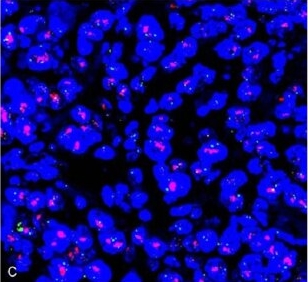}
\small (c) FISH
\end{minipage}

\caption{Different techniques for HER2 detection.}
\label{fig:Imaging_techs}
\end{figure}

Hematoxylin and eosin (H\&E) staining is the standard histopathological technique used for initial cancer screening. Unlike IHC, H\&E is cost-effective, rapid, and independent of antibody selection, making it widely used for cancer diagnosis. H\&E stains nuclei blue to dark purple and cytoplasm in various shades of pink and red, providing high-contrast images for structural analysis. Figure \ref{fig:Imaging_techs} illustrates the three primary techniques used in HER2 detection: H\&E, IHC, and FISH.

Recent advances in deep learning have demonstrated that H\&E images alone can predict HER2 status without IHC \cite{shamai2019artificial}. Studies \cite{gamble2021determining, naik2020deep, bychkov2021deep, rawat2020deep, levy2020preliminary} have used convolutional neural networks (CNNs) and transformers to classify HER2 expression directly from H\&E images. However, a direct H\&E-based classification lacks interpretability, which limits its adoption in clinical practice. Instead, a more promising approach is to translate H\&E images into IHC-like images, allowing human pathologist validation and dataset augmentation.

Liu \textit{et al.} \cite{liu2022bci} pioneered this approach using a GAN-based pyramid pix2pix model, training it on their BCI dataset. While pyramid pix2pix improved image translation, it struggled to generate realistic IHC 3+ images, suffering from mode collapse, where the model fails to capture diverse HER2-positive morphologies. This limitation undermines the clinical utility of the generated IHC images, particularly for HER2-positive cases where a precise classification is required.

This work aims to address the shortcomings of existing image translation models and proposes a novel GAN-based framework with an enhanced loss function. By modifying the loss function, we mitigate mode collapse problem, a fundamental limitation in generative adversarial networks (GANs), and introduce a novel variance-based penalty that enforces structural diversity in generated images. Specifically, our work seeks to:

\begin{enumerate}
    \item Improve the translation of H\&E images into high-fidelity IHC images, particularly for HER2-positive (IHC 3+) cases, which remain a challenge for existing models.
    \item Mitigate mode collapse by introducing a variance-based loss function, ensuring greater diversity in generated images.
    \item Evaluate the generalizability of our model beyond medical imaging by testing it on generic image translation tasks.
\end{enumerate}

To achieve these objectives, we introduce a variance penalty term into the loss function of the pyramid pix2pix model, enforcing structural diversity and fidelity in IHC image generation. Our model is rigorously evaluated using the BCI dataset for HER2 classification and the facades dataset \cite{isola2017image} for general image translation, demonstrating superior performance over existing approaches.

The remainder of this paper is organized as follows. Section 2 reviews the related literature on breast cancer detection and image-to-image translation techniques. Section 3 presents the proposed methodology. Section 4 describes the experimental setup, including dataset analysis and the evaluation metrics employed. Sections 5 and 6 report and discuss the comparative results. Finally, Section 7 concludes the study and outlines potential directions for future research.

\textbf{Source Code:} 
The algorithmic implementation, developed in \texttt{Python}, is openly available for replication and further development at our GitHub repository: \href{https://github.com/SaraRehmat/HE2IHC}{\texttt{https://github.com/SaraRehmat/HE2IHC}}.

\section{Literature Review}

Breast cancer is among the most prevalent malignancies worldwide, requiring advanced techniques for early detection and classification. Significant research efforts have been devoted to automating breast cancer diagnosis using a combination of image processing, machine learning, and deep learning techniques. Traditional image processing methods have focused on nuclei segmentation and feature extraction to differentiate between malignant and benign tumors \cite{xu2014efficient}. With the advent of machine learning, algorithms such as support vector machines (SVMs) and random forests have demonstrated high efficacy in breast cancer classification \cite{wang2018support,kumar2020prediction,lahoura2021cloud}. More recently, deep learning architectures, particularly convolutional neural networks (CNNs), have significantly enhanced histopathological image classification by automatically learning hierarchical representations \cite{hirra2021breast,jiang2019breast,khan2019novel,xie2019deep}.

The recent emergence of Vision Transformers (ViTs) \cite{dosovitskiy2020image} and their variants \cite{touvron2021training,liu2021swin} has further revolutionized the field. Unlike CNNs, which rely on localized receptive fields, ViTs utilize self-attention mechanisms to model long-range dependencies in images, leading to improved feature extraction and classification. Several studies have successfully deployed ViTs for breast cancer detection \cite{alotaibi2023vit,tummala2022breast,shao2021transmil,gul2022histopathological,wang2022semi}, demonstrating superior performance over traditional CNNs.

An important aspect of automated breast cancer diagnosis is the availability of high-quality datasets. The \textit{BreakHis} dataset \cite{spanhol2015dataset} is widely used in breast cancer classification, consisting of 9,109 microscopic images of benign and malignant tumors obtained at different magnifications. Additionally, the BreCaHAD (Breast Cancer Histopathological Annotation and Diagnosis) dataset \cite{aksac2019brecahad} and the BACH (Breast Cancer Histology Challenge) dataset \cite{aresta2019bach} provide rich sources of labeled histopathological images for benchmarking various computational models.

Beyond morphological classification, biomarker analysis plays a pivotal role in determining treatment strategies for breast cancer patients. Immunohistochemistry (IHC) is a widely used method to assess biomarkers such as estrogen receptors (ER), progesterone receptors (PR), and human epidermal growth factor receptor 2 (HER2), which guide targeted therapies. Recent studies \cite{shamai2019artificial,naik2020deep,gamble2021determining} have demonstrated that machine learning algorithms can predict biomarker expression directly from hematoxylin and eosin (H\&E)-stained images, highlighting a strong correlation between morphological structures in H\&E images and biochemical properties observed in IHC images. This has motivated researchers to explore the translation of H\&E images into IHC images as a means to facilitate biomarker assessment while reducing costs and dependency on specialized laboratory techniques.

Image-to-image translation is a critical area of research in computational pathology, particularly for the transformation of H\&E-stained images into IHC images. Liu \textit{et al.} \cite{liu2022bci} pioneered this effort by introducing a supervised deep learning framework for H\&E to IHC image translation. Broadly, image translation methods can be categorized into unsupervised and supervised approaches. Unsupervised translation models, such as CycleGAN \cite{zhu2017unpaired}, DualGAN \cite{yi2017dualgan}, and UNIT (Unsupervised Image-to-Image Translation Network) \cite{liu2017unsupervised}, leverage cycle consistency losses to map images between domains without paired training data. TC-CycleGAN \cite{huang2023tc} proposed an improved CycleGAN-based unpaired translation framework to translate H\&E images into IHC images. The method introduced self-attention modules in both the generator and discriminator to better capture long-range dependencies, and a Gabor-based texture loss to preserve fine texture details.However, this and other unsupervised translation methods often struggle with preserving fine-grained histopathological details. In contrast, supervised models utilize paired datasets to learn a direct mapping between input and target images, resulting in more precise translations.

Among the supervised image translation models, the pix2pix framework \cite{isola2017image} introduced conditional generative adversarial networks (cGANs) for image-to-image translation. Pix2pixHD \cite{wang2018high} extended this approach to generate high-resolution images using coarse-to-fine generators and multi-scale discriminators. Building on these advancements, Liu \textit{et al.} proposed pyramid pix2pix, an enhanced version of pix2pix tailored for translating H\&E images into IHC images \cite{liu2022bci}. Their key contribution was the creation of the BCI dataset, a large-scale paired dataset of H\&E and IHC images. While pyramid pix2pix improved translation quality for low HER2 expression levels (IHC 0/1+/2+), it failed to generate realistic IHC 3+ images, suffering from mode collapse, a common limitation in generative adversarial networks (GANs). \cite{Liu2023MGGANAM} suggested using multi-generator in their GAN-based architecture. The multi-generator has two U-shaped networks at different scales designed to capture low-frequency components like structure and global layout and high frequency components like texture and staining details. They also modified the loss function of pix2pix by adding a cross-entropy regularization term. But their method also struggled to accurately translate IHC 3+ samples and identify dark regions with high HER2 expression. IHC-GAN \cite{saad2025automatic} used a similar dual-scale generator. The framework first employed a MobileNetV3 classifier to predict the HER2 expression level (0, 1+, 2+, or 3+) from the input H\&E image and extract discriminative feature embeddings. The features were then incorporated into a dual-scale generator through Adaptive Instance Normalization (AdaIN), enabling the generator to modulate staining style according to the predicted HER2 stage. The generator consisted of two coupled encoder–decoder branches with inverted residual blocks to reduce parameters. The model used a composite loss that combined adversarial loss, L1 reconstruction loss, perceptual (VGG-based) loss, and feature-matching loss to balance visual fidelity and structural consistency. The pipeline combined a classifier, dual-scale generator, AdaIN conditioning, and multi-scale discriminator, resulting in a structurally complex system that could lead to overfitting in limited medical datasets. 

Addressing mode collapse remains a critical challenge in GAN-based image translation. Several techniques have been proposed to mitigate this issue, including Wasserstein GANs (WGANs) \cite{adler2018banach}, spectral normalization \cite{miyato2018spectral}, mode regularization \cite{che2016mode}, and self-attention mechanisms \cite{zhang2019self}. However, none of these strategies have been explicitly applied to the translation of H\&E images into IHC images. This gap underscores the need for novel loss functions that penalize mode collapse while maintaining fidelity and structural diversity in generated IHC images.

Given these limitations, our study introduces an enhanced loss function that explicitly penalizes low-diversity image generation by incorporating a variance-based regularization term. By addressing mode collapse, our approach improves the quality and diversity of IHC image generation, thereby advancing computer-aided HER2 breast cancer diagnostics. The proposed methodology is not only specific to histopathological image translation but also demonstrates promising results for generic image-to-image translation tasks, extending its applicability beyond biomedical imaging.

\section{Proposed Methodology}

The transformation of hematoxylin and eosin (H\&E) stained images into Immunohistochemistry (IHC) stained images is a fundamental challenge in digital pathology. Accurate translation between these imaging modalities is essential for both diagnostic and prognostic applications. However, existing generative models, such as pyramid pix2pix \cite{liu2022bci}, often suffer from mode collapse, leading to reduced variability in generated images and suboptimal generalization across various histopathological samples, especially in translating HER2-positive (IHC 3+) images. 

\begin{figure*}[htbp]
    \centering
    \includegraphics[width=\textwidth]{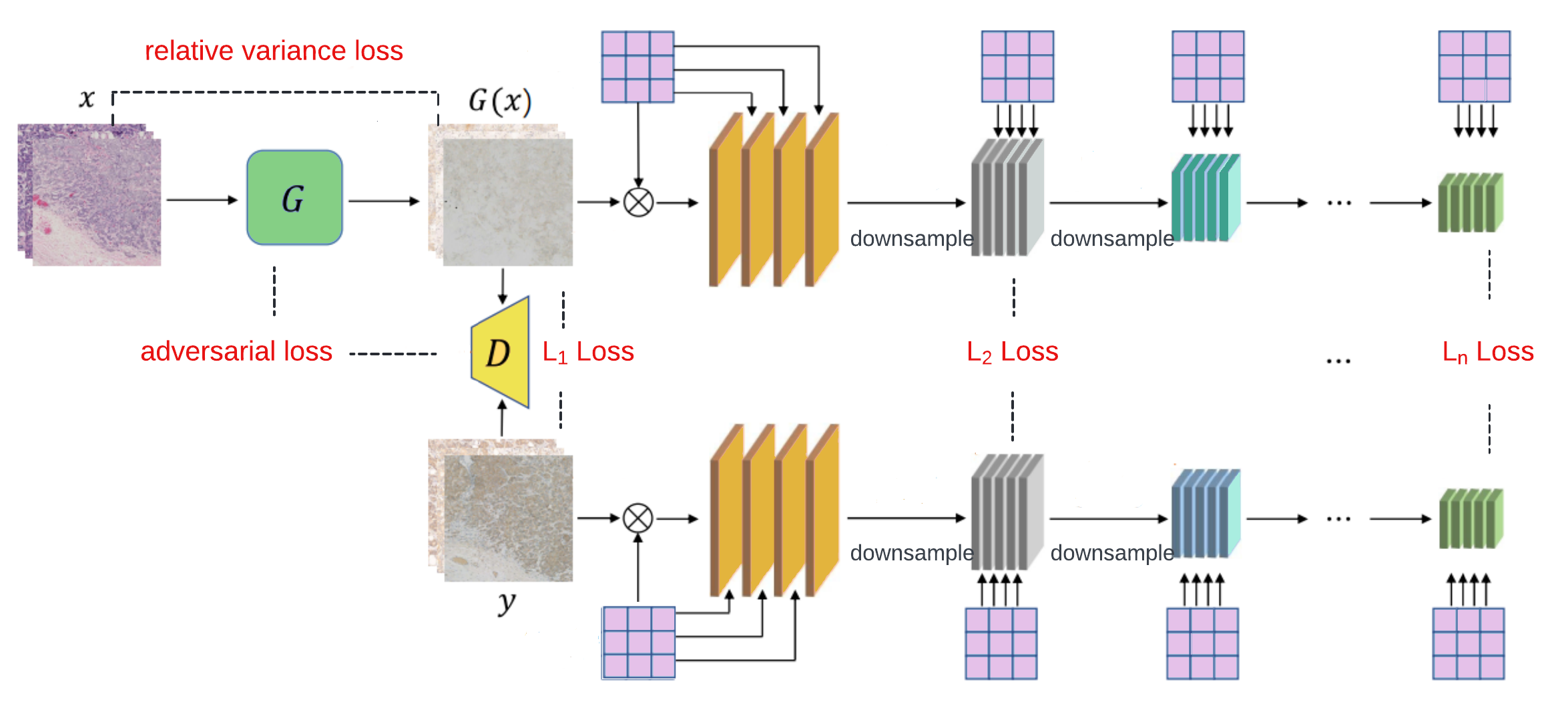}
    \caption{The framework of the proposed architecture for generating high-fidelty IHC images from H\&E images. The relative variance between input images and generated images in a batch is incorporated into the overall loss of pyramid pix2pix \cite{liu2022bci}.}
    \label{fig:ours_arch}
\end{figure*}

To overcome these challenges, we propose an enhancement to the loss function of pyramid pix2pix framework by introducing a variance difference regularization term. The general framework of our approach is illustrated in Figure \ref{fig:ours_arch}. The proposed modification encourages diversity in generated images, while maintaining fidelity to the original IHC image distributions. By penalizing lower variance in generated images relative to the input images, the model's learning is driven to generate more diverse and realistic representations, thereby reducing mode collapse and improving translation accuracy.

\subsection{Pyramid pix2pix Framework}
Pyramid pix2pix is an advanced image-to-image translation model that introduces multi-scale loss computation to mitigate misalignment issues present in the BCI dataset. The original objective function of pyramid pix2pix is defined as:

\begin{equation}
    {G}^{\ast} = \arg \min_{G}\max_{D} \mathcal{L}_{\text{cGAN}}(G,D ) + \lambda \mathcal{L}_1 +  \mathcal{L}_{\text{multiscale}} 
    \label{eq:pix2pixPyramid}
\end{equation}

where \( \mathcal{L}_{\text{cGAN}} \) represents the adversarial loss, \( \mathcal{L}_1 \) is the pixel-wise loss, and \( \mathcal{L}_{\text{multiscale}} \) accounts for structural consistency across scales.

The adversarial loss \( \mathcal{L}_{\text{cGAN}} \) is given by:

\begin{equation}
\mathcal{L}_{\text{cGAN}}(G, D) = \mathbb{E}[\log D(x, y)] + \mathbb{E}[\log (1 - D(x,G(x, z)))]
\label{eq:adv}
\end{equation}

where \( z \) denotes a random noise vector, \( x \) is the H\&E image, \( y \) is the corresponding IHC image, and \( G(x, z) \) is the generated output. The discriminator \( D \) aims to classify real image pairs correctly, while the generator \( G \) seeks to fool \( D \) by generating realistic images.

The L1 loss enforces pixel-wise similarity and is expressed as:

\begin{equation}
\mathcal{L}_1(G) = \mathbb{E}_{x,y,z} \left[ \left| y - G(x,z) \right| \right]
\label{eq:L1}
\end{equation}

Additionally, the multi-scale loss \( \mathcal{L}_{\text{multiscale}} \) is computed as:

\begin{equation}
 \mathcal{L}_{\text{multiscale}}(G) = \sum_{i} \lambda_i S_i 
 \label{eq:multi}
\end{equation}

where \( S_i \) represents structural differences computed at different scales:

\begin{equation}
S_i(G) = \mathbb{E}_{x,y,z} \left[ \left|F_i(y) - F_i(G(x,z)) \right| \right]
\label{eq:s_i}
\end{equation}

\subsection{Variance-Based Loss for Mode Collapse Mitigation}
One of the key challenges in generative models is mode collapse, where the generator produces limited variations of output images. To address this issue, we introduce a variance-based regularization term that penalizes low variance in generated images. The variance computation is performed across a batch of \( n \) images with \( C \) channels, height \( H \), and width \( W \).

First, the mean pixel value at each channel and spatial location is computed across the batch as:

\begin{equation}
\mu_f(c,i,j) = \frac{1}{n} \sum_{k=1}^{n} f_{k,c,i,j},
\qquad
\mu_g(c,i,j) = \frac{1}{n} \sum_{k=1}^{n} G(f)_{k,c,i,j}
\end{equation}

The batch-wise variance at each channel and spatial position is then:

\begin{equation}
\sigma_f^2(c,i,j) = \frac{1}{n} \sum_{k=1}^{n} \left(f_{k,c,i,j}-\mu_f(c,i,j)
\right)^2
\end{equation}

\begin{equation}
\sigma_g^2(c,i,j)
\frac{1}{n}
\sum_{k=1}^{n}
\left(
G(f)_{k,c,i,j}-\mu_g(c,i,j)
\right)^2
\end{equation}

Next, the absolute variance difference is computed at each pixel location:

\begin{equation}
\Delta \sigma^2(c,i,j)=\left|\sigma_f^2(c,i,j)-
\sigma_g^2(c,i,j)
\right|
\end{equation}

Finally, the proposed variance regularization loss is obtained by averaging these absolute differences over all channels and spatial locations:

\begin{equation}
\mathcal{L}_{var}=\frac{1}{C \cdot H \cdot W}
\sum_{c=1}^{C}
\sum_{i=1}^{H}
\sum_{j=1}^{W}
\Delta \sigma^2(c,i,j)
\label{eq:rel_var}
\end{equation}

This variance regularization term ensures that the generator maintains diversity in generated outputs. Figure \ref{fig:rel_var} illustrates the tensor operations involved in computing variance.

\subsection{Final Objective Function}
By incorporating the variance-based regularization term defined in equation \ref{eq:rel_var} into the pyramid pix2pix loss function, we obtain the following final objective function:

\begin{equation}
 {G}^{\ast} = \arg \min_{G}\max_{D} \mathcal{L}_{cGAN}(G,D ) + \lambda \mathcal{L}_1 + \mathcal{L}_{\text{multiscale}}+\mathcal{L}_{var} 
 \label{eq:ours} 
\end{equation}

The inclusion of \( \mathcal{L}_{var} \) mitigates mode collapse by encouraging variance in generated images while maintaining structural consistency. 
\begin{figure*}[t]
\centering
\begin{tikzpicture}[
    tensor/.style={
      draw, fill=blue!10,
      minimum height=1.05cm, minimum width=2.7cm,
      align=center, font=\small, text width=2.7cm
    },
    box/.style={
      draw, rounded corners, fill=orange!20,
      minimum height=1.05cm, minimum width=3.2cm,
      align=center, font=\small, text width=3.2cm
    },
    arrow/.style={-{Latex[scale=1.2]}, thick},
    node distance=1.4cm and 2.2cm
]

\node[tensor] (f) {$f$: Real Images\\$n \times C \times H \times W$};
\node[tensor, below=of f] (g) {$g$: Generated Images\\$n \times C \times H \times W$};

\node[box, right=of f] (sigma2f) {$\sigma^2(f)$\\Variance per pixel (real)};
\node[box, right=of g] (sigma2g) {$\sigma^2(g)$\\Variance per pixel (gen)};

\node[box, right=2.4cm of $(sigma2f)!0.5!(sigma2g)$] (absdiff)
{$|\sigma^2(f) - \sigma^2(g)|$\\Pixel-wise absolute difference};

\node[box, right=of absdiff] (avg)
{$\mathcal{L}_{var}$\\Average over all pixels};

\draw[arrow] (f) -- (sigma2f);
\draw[arrow] (g) -- (sigma2g);
\draw[arrow] (sigma2f) -- (absdiff);
\draw[arrow] (sigma2g) -- (absdiff);
\draw[arrow] (absdiff) -- (avg);


\end{tikzpicture}
\caption{Computation of the proposed variance loss. After computing batch-wise pixel variance for real and generated tensors, the absolute difference is calculated at each pixel location, followed by averaging to obtain the final variance loss.}
\label{fig:rel_var}
\end{figure*}

\section{Experimental Setup}

To evaluate the effectiveness of our proposed variance-based loss enhancement in pyramid pix2pix, we conduct experiments on a publicly available dataset \cite{liu2022bci} of paired hematoxylin and eosin (H\&E) and Immunohistochemistry (IHC) stained images. The dataset comprises high-resolution histopathological images from multiple tissue samples, ensuring diversity in staining patterns and structural variations. We preprocess all images by normalizing intensity values and resizing them to a fixed resolution suitable for deep learning models. The training and validation splits are maintained at an 80:20 ratio to ensure a balanced evaluation. Our experiments are implemented using PyTorch and executed on Google Colab, leveraging its Tesla T4 GPU acceleration. The model is trained with a batch size of 8, an initial learning rate of 0.0002, and the Adam optimizer to ensure stable convergence.

\subsection{Dataset Analysis}

\begin{figure}[htbp]
\centering
\includegraphics[width=0.75\linewidth]{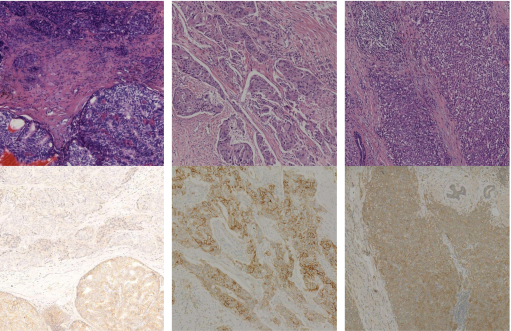}
\caption{Representative image patches from the BCI dataset \cite{liu2022bci}, showcasing hematoxylin and eosin (H\&E) stained images alongside their corresponding immunohistochemistry (IHC) images. The dataset provides paired samples for deep learning-based image translation and biomarker assessment.}
\label{fig:HE2IHC}
\end{figure}

The BCI dataset curated by \cite{liu2022bci} serves as a comprehensive resource specifically designed for histopathological image translation tasks. It comprises 4873 paired image patches extracted from whole slide images of 51 patients, covering a broad spectrum of histological variations. The dataset includes all four categories of HER2 expression in IHC images: 0, 1+, 2+, and 3+. Each patch has a high resolution of 1024 x 1024 pixels, preserving fine-grained histopathological details necessary for accurate image-to-image translation. The paired nature of the dataset makes it particularly suitable for supervised learning approaches, where each H\&E-stained image is directly associated with a corresponding IHC counterpart, as illustrated in Figure  \ref{fig:HE2IHC}.

\begin{figure}[htbp]
\centering
\begin{tikzpicture}
\pie[
    radius=2.5,  
    text=legend,  
    color={red!30, green!30, blue!30, yellow!30},  
    sum=100  
    ]{
    5.14/IHC 0 (5.14\%),
    24.7/IHC 1+ (24.7\%),
    45.9/IHC 2+ (45.9\%),
    24.3/IHC 3+ (24.3\%)
    }
\end{tikzpicture}
\caption{Distribution of HER2 expression levels in the BCI dataset. The dataset is predominantly composed of IHC 2+ images, reflecting clinical scenarios where further fluorescence in situ hybridization (FISH) confirmation is required.}
\label{fig:IHC_distribution}
\end{figure}

The dataset exhibits a non-uniform distribution of IHC image patches across the four HER2 expression levels, reflecting real-world prevalence and variability in HER2 staining patterns. As depicted in Figure \ref{fig:IHC_distribution}, a higher proportion of images correspond to IHC 2+ expression, which represents a critical threshold category that requires additional FISH confirmation in clinical settings. The diversity of samples within the dataset is helpful in training models that can generalize effectively across different staining intensities and cellular morphologies. The variation in sample representation highlights the necessity of developing robust translation models capable of accurately generating high-quality IHC images, even for underrepresented categories.

\begin{figure}[htbp]
\centering
\begin{tikzpicture}
\begin{axis}[
    ybar,
    symbolic x coords={IHC 0, IHC 1+, IHC 2+, IHC 3+},
    xtick=data,
    ylabel={Average Color Means (RGB)},
    legend style={at={(0.5,-0.15)}, anchor=north,legend columns=-1},
    ymin=150, ymax=210,
    width=9cm, height=7cm,
    bar width=10pt,
    enlarge x limits=0.2,
    nodes near coords,
     nodes near coords style={font=\scriptsize}, 
]
\addplot coordinates {(IHC 0,198) (IHC 1+,176) (IHC 2+,173) (IHC 3+,160)};
\addplot coordinates {(IHC 0,191) (IHC 1+,167) (IHC 2+,165) (IHC 3+,154)};
\addplot coordinates {(IHC 0,202) (IHC 1+,183) (IHC 2+,180) (IHC 3+,170)};

\legend{R Mean, G Mean, B Mean}
\end{axis}
\end{tikzpicture}
\caption{Mean RGB color values for different HER2 expression levels in the BCI dataset. Distinct color variations highlight staining intensity differences across IHC categories.}
    \label{fig:color_means}
\end{figure}

\begin{figure}[htbp]
\centering
\begin{tikzpicture}
\begin{axis}[
    ybar,
    symbolic x coords={IHC 0, IHC 1+, IHC 2+, IHC 3+},
    xtick=data,
    ylabel={Average Sharpness},
    ymin=100, ymax=200,
    width=9cm, height=7cm,
    bar width=15pt,
    enlarge x limits=0.2,
    nodes near coords,
]
\addplot coordinates {(IHC 0,146.15) (IHC 1+,118.81) (IHC 2+,132.05) (IHC 3+,182.95)};
\end{axis}
\end{tikzpicture}
\caption{Average sharpness values for different HER2 expression levels. The variability in sharpness reflects structural and textural differences among IHC categories, influencing diagnostic precision.}
\label{fig:sharpness}
\end{figure}

\begin{figure}[t]
\centering
\begin{tikzpicture}
\begin{axis}[
    ybar,
    symbolic x coords={IHC 0, IHC 1+, IHC 2+, IHC 3+},
    xtick=data,
    ylabel={Average Dynamic Range},
    ymin=200, ymax=235,
    width=9cm, height=7cm,
    bar width=15pt,
    enlarge x limits=0.2,
    nodes near coords,
]
\addplot coordinates {(IHC 0,227.93) (IHC 1+,216.20) (IHC 2+,208.24) (IHC 3+,205.61)};
\end{axis}
\end{tikzpicture}
\caption{Comparison of dynamic range values across HER2 expression levels. The decreasing trend in dynamic range correlates with increased HER2 expression, impacting the contrast in generated IHC images.}
\label{fig:dynamic_range}
\end{figure}

A detailed analysis of the dataset reveals significant variability in terms of color distribution, sharpness, and dynamic range across different HER2 expression levels. Figure \ref{fig:color_means} illustrates differences in the average color means, highlighting distinct staining properties among different categories. Similarly, Figure \ref{fig:sharpness} and Figure \ref{fig:dynamic_range} provide insights into the texture and contrast variations between H\&E and IHC images. These differences present challenges for existing generative models, particularly in maintaining structural integrity while ensuring realistic color translation. Addressing these variations remains essential for improving the fidelity of generated IHC images and enhancing their applicability in clinical diagnostic workflows.

\subsection{Evaluation Metrics}

To assess the quality of the generated immunohistochemistry (IHC) images from hematoxylin and eosin (H\&E) stained images, we employ three widely used evaluation metrics: Peak Signal-to-Noise Ratio (PSNR), Structural Similarity Index (SSIM) \cite{wang2004image}, and Fr\'echet Inception Distance (FID) \cite{heusel2017gans}. Each of these metrics provides a complementary perspective on the accuracy and perceptual quality of the translated images.

\subsubsection{Peak Signal-to-Noise Ratio (PSNR)}
PSNR is a full-reference metric that quantifies the pixel-wise fidelity of the generated image with respect to the ground truth image. It is derived from the Mean Squared Error (MSE), which measures the pixel-wise difference between the original and generated images. Given an original image $ I $ and a reconstructed image $ K $ of size $ m \times n $, the MSE is computed as:

\begin{equation}
\text{MSE} = \frac{1}{mn} \sum_{i=0}^{m-1} \sum_{j=0}^{n-1} [I(i, j) - K(i, j)]^2.
\end{equation}

Using MSE, PSNR is then calculated as:

\begin{equation}
\text{PSNR} = 10 \log_{10} \left( \frac{L^2}{\text{MSE}} \right),
\end{equation}

where $ L $ is the maximum possible pixel value of the image (e.g., 255 for 8-bit images). Higher PSNR values indicate greater similarity between the generated and ground truth images, signifying lower reconstruction error. In the context of HE to IHC translation, PSNR provides an objective measure of how closely the pixel intensities in the generated IHC image match the real IHC counterpart. However, since PSNR does not account for perceptual quality and structural differences, it is best used in conjunction with other metrics.

\subsubsection{Structural Similarity Index (SSIM)}
SSIM is designed to model human visual perception by evaluating luminance, contrast, and structural similarity between the generated and ground truth images. Unlike PSNR, which only considers absolute differences, SSIM assesses structural distortions, which are crucial in medical image translation tasks where fine morphological details must be preserved. SSIM is computed as:

\begin{equation}
\text{SSIM}(x, y) = \frac{(2 \mu_x \mu_y + C_1)(2 \sigma_{xy} + C_2)}{(\mu_x^2 + \mu_y^2 + C_1)(\sigma_x^2 + \sigma_y^2 + C_2)},
\end{equation}

where:
\begin{align*}
\mu_x & \text{ is the mean intensity of image } x, \\
\mu_y & \text{ is the mean intensity of image } y, \\
\sigma_x^2 & \text{ is the variance of image } x, \\
\sigma_y^2 & \text{ is the variance of image } y, \\
\sigma_{xy} & \text{ is the covariance between images } x \text{ and } y, \\
C_1 & = (K_1 L)^2, \\
C_2 & = (K_2 L)^2.
\end{align*}

Higher SSIM values (closer to 1) indicate a greater structural resemblance between the two images. Since medical images rely heavily on structural integrity for clinical interpretation, SSIM is particularly valuable in evaluating H\&E to IHC translations, ensuring that spatial structures are preserved despite color and texture transformations.

\subsubsection{Fr\'echet Inception Distance (FID)}
FID assesses the quality and diversity of the generated images by comparing the statistical distribution of deep features extracted from a pre-trained Inception v3 model \cite{szegedy2016rethinking}. Unlike PSNR and SSIM, which evaluate pixel-wise differences, FID provides a perceptual similarity measure at a higher abstraction level. It is defined as:

\begin{equation}
\text{FID}(P_r, P_g) = || \mu_r - \mu_g ||^2 + \text{Tr}(\Sigma_r + \Sigma_g - 2(\Sigma_r \Sigma_g)^{1/2}),
\end{equation}

where:
\begin{itemize}
    \item $ \mu_r $ and $ \mu_g $ are the means of the feature activations for real and generated images, respectively.
    \item $ \Sigma_r $ and $ \Sigma_g $ are the covariance matrices of the feature activations for real and generated images, respectively.
    \item $ \text{Tr} $ denotes the trace of a matrix.
\end{itemize}

Lower FID values indicate greater similarity between the distributions of real and generated images, reflecting improved realism and diversity in the generated IHC images. In the context of HE to IHC translation, FID ensures that the generated images not only resemble real IHC images in appearance but also capture the statistical properties of real IHC data, which is crucial for clinical applicability.

\subsubsection{Summary of Metrics}
A high-quality HE to IHC image translation model should produce images with:
\begin{itemize}
    \item Higher PSNR values, indicating lower pixel-wise reconstruction error.
    \item Higher SSIM values, ensuring structural similarity with the ground truth.
    \item Lower FID values, demonstrating better perceptual realism and diversity.
\end{itemize}
Using these complementary metrics, we provide a comprehensive evaluation of our model's ability to generate high-fidelity IHC images from H\&E-stained input images, ensuring both visual and statistical alignment with real IHC data.

\subsection{Experimental Framework}

The BCI dataset \cite{liu2022bci} we used consists of 4873 paired hematoxylin and eosin (H\&E) and HER2 Immunohistochemistry (IHC) stained images. The IHC images are further classified into four categories based on staining intensity and membrane completeness, which help determine HER2 expression levels in tumor cells. IHC 0 is considered negative, showing no staining or faint, incomplete membrane staining in less than 10\% of tumor cells. IHC 1+ is also negative but presents weak and incomplete membrane staining in more than 10\% of tumor cells. IHC 2+ is considered equivocal, displaying weak to moderate complete membrane staining in over 10\% of tumor cells, often requiring fluorescence in situ hybridization (FISH) for confirmation. IHC 3+, classified as HER2-positive, exhibits strong, circumferential membrane staining in more than 10\% of tumor cells, confirming HER2 overexpression without the need for additional testing.

The training dataset comprises images exhibiting diverse characteristics based on HER2 concentration levels, as reflected by variations in statistical properties such as color distribution, sharpness, and dynamic range. These variations are quantitatively illustrated in Figures \ref{fig:color_means}, \ref{fig:sharpness}, and \ref{fig:dynamic_range}. 

To evaluate our proposed method, we compare the performance metrics, \textit{Peak Signal-to-Noise Ratio (PSNR)}, \textit{Structural Similarity Index (SSIM)}, and \textit{Fréchet Inception Distance (FID)}, against two baseline methods: \textit{pix2pix} \cite{isola2017image} and \textit{pyramid pix2pix} \cite{liu2022bci}. To optimize computational efficiency, input images are cropped before being fed into the generator. 

The generator in our proposed framework utilizes a \textit{ResNet-9blocks} architecture, while the discriminator follows the default \textit{PatchGAN} configuration. Training is conducted with a batch size of \textit{8} for \textit{100 epochs}. The optimization is performed using the \textit{Adam} optimizer, with an initial learning rate of \textit{0.0002} for the first \textit{50 epochs}, which then linearly decays to zero over the remaining epochs. The momentum parameter for Adam is set to \textit{0.5}. The generator is trained using the loss function defined in Equation \ref{eq:ours}, incorporating our proposed variance-based loss term.

\section{Results}

To comprehensively assess the effectiveness of our proposed model, we evaluated its performance across multiple classes of images, focusing on its ability to generate high-fidelity immunohistochemistry (IHC) images from hematoxylin and eosin (H\&E) stained images. Our results are presented in comparison with state-of-the-art models, including pix2pix \cite{isola2017image} and pyramid pix2pix \cite{liu2022bci}, using the Breast Cancer Immunohistochemistry (BCI) dataset.

\subsection{Comparison with Existing Methods on BCI Dataset}

To assess the effectiveness of our proposed model in translating hematoxylin and eosin (H\&E) stained images into high-fidelity immunohistochemistry (IHC) images, particularly for HER2-positive (IHC 3+) cases, we performed a detailed quantitative and qualitative analysis.

\subsubsection{Quantitative Evaluation on Test Set}

 Figure \ref{fig:classwise_metrics} illustrates the peak signal-to-noise ratio (PSNR), the structural similarity index (SSIM), and the Fréchet inception distance (FID) for different levels of expression of HER2. Our model consistently demonstrates superior performance, as evident by higher SSIM, PSNR, and lower FID scores, particularly for IHC 3+ images. This confirms that our method enhances the fidelity of IHC image synthesis, addressing the primary challenge of accurately reconstructing HER2-positive cases, which has remained a limitation of previous models.

\begin{figure}[htbp]
\centering

\begin{minipage}{0.48\linewidth}
\centering
\begin{tikzpicture}
\begin{axis}[
    ybar,
    symbolic x coords={IHC 0, IHC 1+, IHC 2+, IHC 3+},
    xtick=data,
    ymin=18,
    ymax=23,
    bar width=13pt,
    width=\linewidth,
    height=6cm,
    enlarge x limits=0.2,
    legend style={at={(0.5,-0.18)}, anchor=north,legend columns=-1},
    nodes near coords,
    nodes near coords style={font=\tiny},
    xlabel={Image Class},
    ylabel={PSNR},
]
\addplot coordinates {(IHC 0,19.67) (IHC 1+,22.22) (IHC 2+,20.9) (IHC 3+,19.2)};
\addplot coordinates {(IHC 0,19.2) (IHC 1+,21.07) (IHC 2+,20.45) (IHC 3+,19.4)};
\addplot coordinates {(IHC 0,20.48) (IHC 1+,21.65) (IHC 2+,21.34) (IHC 3+,19.5)};

\legend{pix2pix, pyramid pix2pix, ours}
\end{axis}
\end{tikzpicture}

\vspace{1mm}
\small (a) PSNR
\end{minipage}
\hfill
\begin{minipage}{0.48\linewidth}
\centering
\begin{tikzpicture}
\begin{axis}[
    ybar,
    symbolic x coords={IHC 0, IHC 1+, IHC 2+, IHC 3+},
    xtick=data,
    ymin=0.3,
    ymax=0.59,
    bar width=12pt,
    width=\linewidth,
    height=6cm,
    enlarge x limits=0.2,
    legend style={at={(0.5,-0.18)}, anchor=north,legend columns=-1},
    nodes near coords,
    every node near coord/.append style={font=\tiny},
    xlabel={Image Class},
    ylabel={SSIM},
]
\addplot coordinates {(IHC 0,0.52) (IHC 1+,0.47) (IHC 2+,0.44) (IHC 3+,0.39)};
\addplot coordinates {(IHC 0,0.46) (IHC 1+,0.44) (IHC 2+,0.42) (IHC 3+,0.38)};
\addplot coordinates {(IHC 0,0.56) (IHC 1+,0.48) (IHC 2+,0.46) (IHC 3+,0.42)};

\legend{pix2pix, pyramid pix2pix, ours}
\end{axis}
\end{tikzpicture}

\vspace{1mm}
\small (b) SSIM
\end{minipage}

\vspace{3mm}

\begin{minipage}{0.62\linewidth}
\centering
\begin{tikzpicture}
\begin{axis}[
    ybar,
    symbolic x coords={IHC 0, IHC 1+, IHC 2+, IHC 3+},
    xtick=data,
    ymin=380,
    ymax=970,
    bar width=11pt,
    width=\linewidth,
    height=6cm,
    enlarge x limits=0.3,
    legend style={at={(0.5,-0.18)}, anchor=north,legend columns=-1},
    nodes near coords,
    nodes near coords style={
        /pgf/number format/fixed,
        /pgf/number format/precision=0,
        /pgf/number format/1000 sep={}
    },
    every node near coord/.append style={font=\tiny},
    yticklabel style={
        /pgf/number format/precision=0,
        /pgf/number format/fixed zerofill,
        /pgf/number format/1000 sep={}
    },
    xlabel={Image Class},
    ylabel={FID},
]
\addplot coordinates {(IHC 0,914) (IHC 1+,605) (IHC 2+,474) (IHC 3+,733)};
\addplot coordinates {(IHC 0,921) (IHC 1+,643) (IHC 2+,527) (IHC 3+,759)};
\addplot coordinates {(IHC 0,890) (IHC 1+,472) (IHC 2+,395) (IHC 3+,508)};

\legend{pix2pix, pyramid pix2pix, ours}
\end{axis}
\end{tikzpicture}

\vspace{1mm}
\small (c) FID
\end{minipage}

\caption{Class-wise quantitative comparison on the test set for Pix2pix, PyramidPix2pix, and the proposed method across IHC score categories (0, 1+, 2+, 3+). (a) PSNR and (b) SSIM (higher is better) measure pixel-level fidelity and structural similarity, while (c) FID (lower is better) reflects distribution-level realism. The proposed method achieves consistently improved PSNR/SSIM and lower FID across most classes, indicating better reconstruction quality and perceptual realism.}
\label{fig:classwise_metrics}
\end{figure}

The evaluation metrics for our methodology, using the BCI dataset, are presented in comparison with existing approaches, namely pix2pix \cite{isola2017image} and pyramid pix2pix \cite{liu2022bci} in Table \ref{tab:avg_comparison} and Figure \ref{fig:comparison}. Notably, our method achieves the highest PSNR and SSIM scores, indicating superior pixel-wise fidelity and enhanced preservation of structural information, both essential for clinical interpretability. Moreover, the substantial reduction in FID highlights the strong resemblance between real and generated images in the feature space, underscoring the effectiveness of our model in achieving high-quality image translation.

Furthermore, our model outperforms existing models by reporting the best evaluation metrics for all classes of HER2 expression as presented in Tables 
\ref{tab:ihc_0}-\ref{tab:ihc_3_plus} and Figure \ref{fig:comparison}. This consistently superior performance across all classes highlights the robustness and reliability of our method.

All these results indicate that our variance-based loss function plays a crucial role in mitigating mode collapse, resulting in a broader diversity of generated images and ensuring that the synthetic IHC images accurately capture the inherent variability observed in real-world HER2-positive cases. This advancement is critical for reliable computational pathology, where the preservation of fine-grained details in IHC images is paramount for clinical decision-making.

\begin{figure}[htbp]
\centering

\begin{minipage}[t]{0.32\textwidth}
\centering
\begin{tikzpicture}
\begin{axis}[
    ybar,
    ymin=18, ymax=26,
    bar width=0.3cm,
    ylabel={PSNR},
    symbolic x coords={pix2pix, pyramid pix2pix, ours},
    xtick=data,
    nodes near coords,
    nodes near coords style={font=\footnotesize},
    xticklabel style={
        text height=2.5ex, text depth=1ex, anchor=center
    },
    width=\textwidth,
    height=5cm,
    enlarge x limits=0.2,
    ylabel style={font=\footnotesize},
    tick label style={font=\footnotesize},
]
\addplot coordinates {(pix2pix,20.74) (pyramid pix2pix,21.15) (ours,22.16)};
\end{axis}
\end{tikzpicture}

\vspace{2mm}
{\small\textbf{(a)} PSNR}
\end{minipage}
\hfill
\begin{minipage}[t]{0.32\textwidth}
\centering
\begin{tikzpicture}
\begin{axis}[
    ybar,
    ymin=0.3, ymax=0.55,
    bar width=0.3cm,
    ylabel={SSIM},
    symbolic x coords={pix2pix, pyramid pix2pix, ours},
    xtick=data,
    nodes near coords,
    nodes near coords style={font=\footnotesize},
    xticklabel style={
        text height=2.5ex, text depth=1ex, anchor=center
    },
    width=\textwidth,
    height=5cm,
    enlarge x limits=0.2,
    ylabel style={font=\footnotesize},
    tick label style={font=\footnotesize},
]
\addplot coordinates {(pix2pix,0.44) (pyramid pix2pix,0.43) (ours,0.47)};
\end{axis}
\end{tikzpicture}

\vspace{2mm}
{\small\textbf{(b)} SSIM}
\end{minipage}
\hfill
\begin{minipage}[t]{0.32\textwidth}
\centering
\begin{tikzpicture}
\begin{axis}[
    ybar,
    ymin=300, ymax=850,
    bar width=0.3cm,
    ylabel={FID},
    symbolic x coords={pix2pix, pyramid pix2pix, ours},
    xtick=data,
    nodes near coords,
    nodes near coords style={font=\footnotesize},
    xticklabel style={
        text height=2.5ex, text depth=1ex, anchor=center
    },
    width=\textwidth,
    height=5cm,
    enlarge x limits=0.2,
    ylabel style={font=\footnotesize},
    tick label style={font=\footnotesize},
]
\addplot coordinates {(pix2pix,472.6) (pyramid pix2pix,516.75) (ours,346.37)};
\end{axis}
\end{tikzpicture}

\vspace{2mm}
{\small\textbf{(c)} FID}
\end{minipage}

\caption{Comparison of image quality metrics across models. (a) PSNR, (b) SSIM, and (c) FID comparison between pix2pix, pyramid pix2pix, and the proposed method for the entire dataset representing all classes of HER2 expression.}
\label{fig:comparison}
\end{figure}

\subsubsection{Qualitative Comparison}

\begin{figure*}[htbp]
\centering

\begin{minipage}[t]{0.48\textwidth}
\centering

\begin{tabular}{>{\centering\arraybackslash}m{1.3cm}
                >{\centering\arraybackslash}m{1.3cm}
                >{\centering\arraybackslash}m{1.3cm}
                >{\centering\arraybackslash}m{1.3cm}
                >{\centering\arraybackslash}m{1.3cm}}
\small H\&E & \small Real IHC & \small pix2pix & \small pyramid pix2pix & \small ours\\
\includegraphics[width=1.5cm]{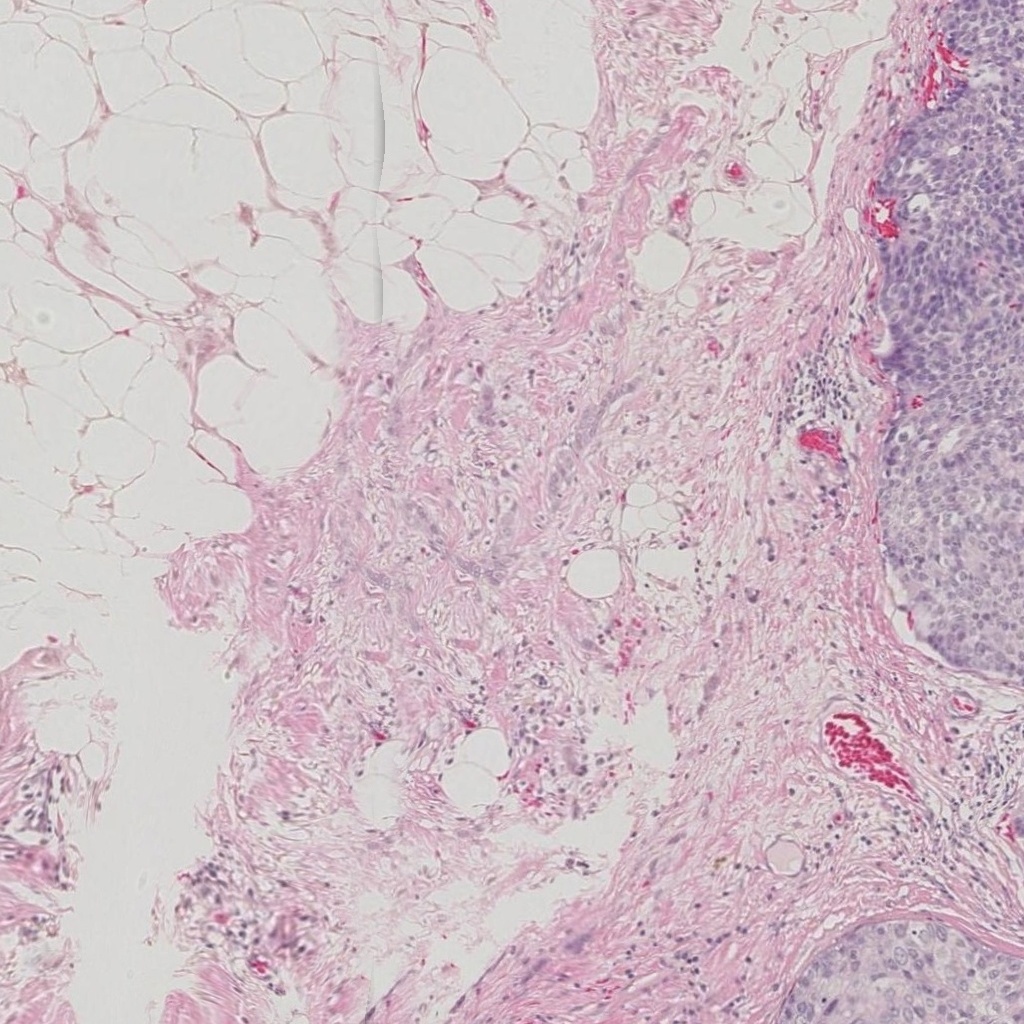} &
\includegraphics[width=1.5cm]{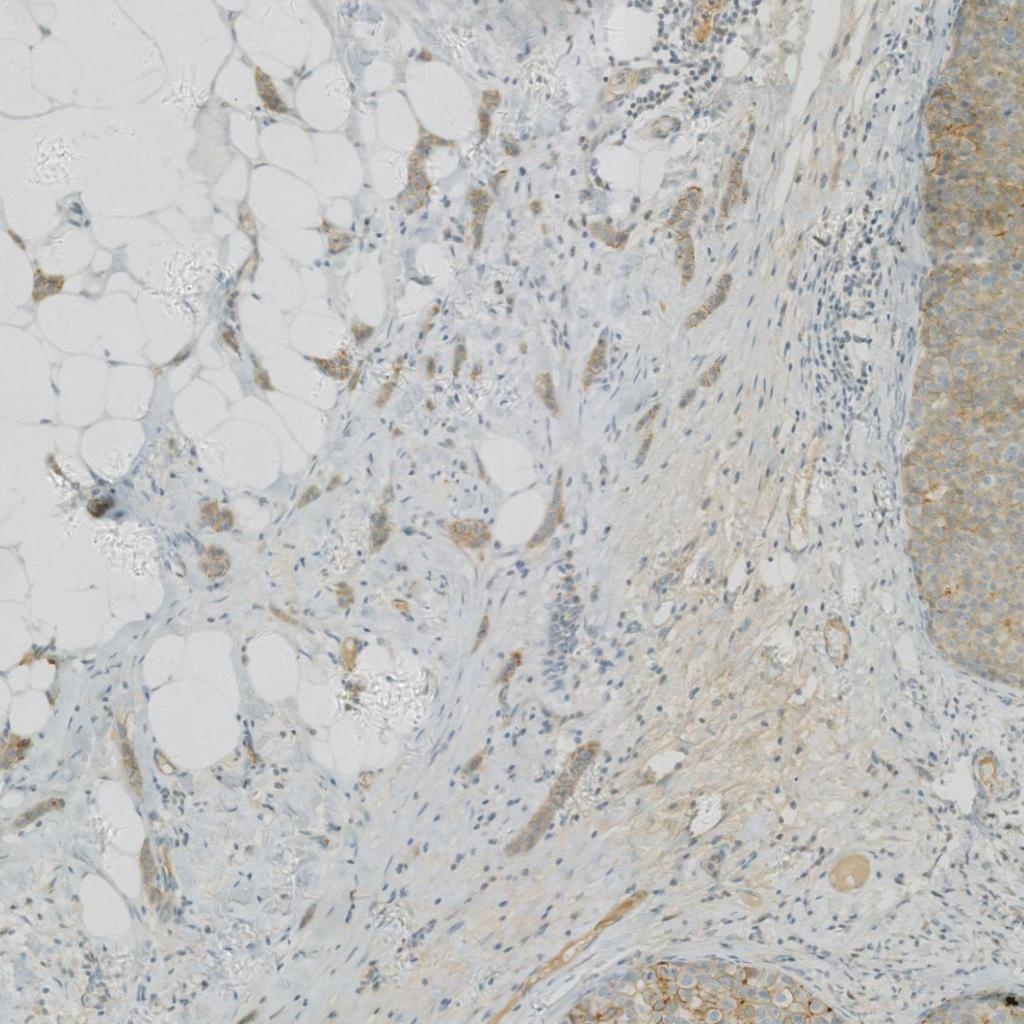} &
\includegraphics[width=1.5cm]{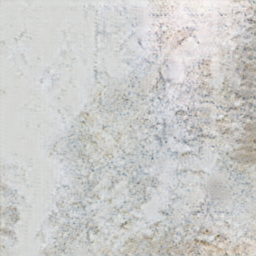} &
\includegraphics[width=1.5cm]{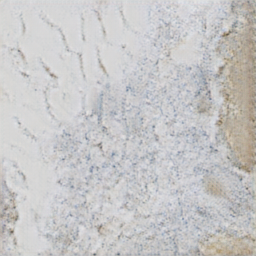} &
\includegraphics[width=1.5cm]{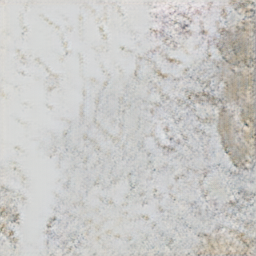} \\
\includegraphics[width=1.5cm]{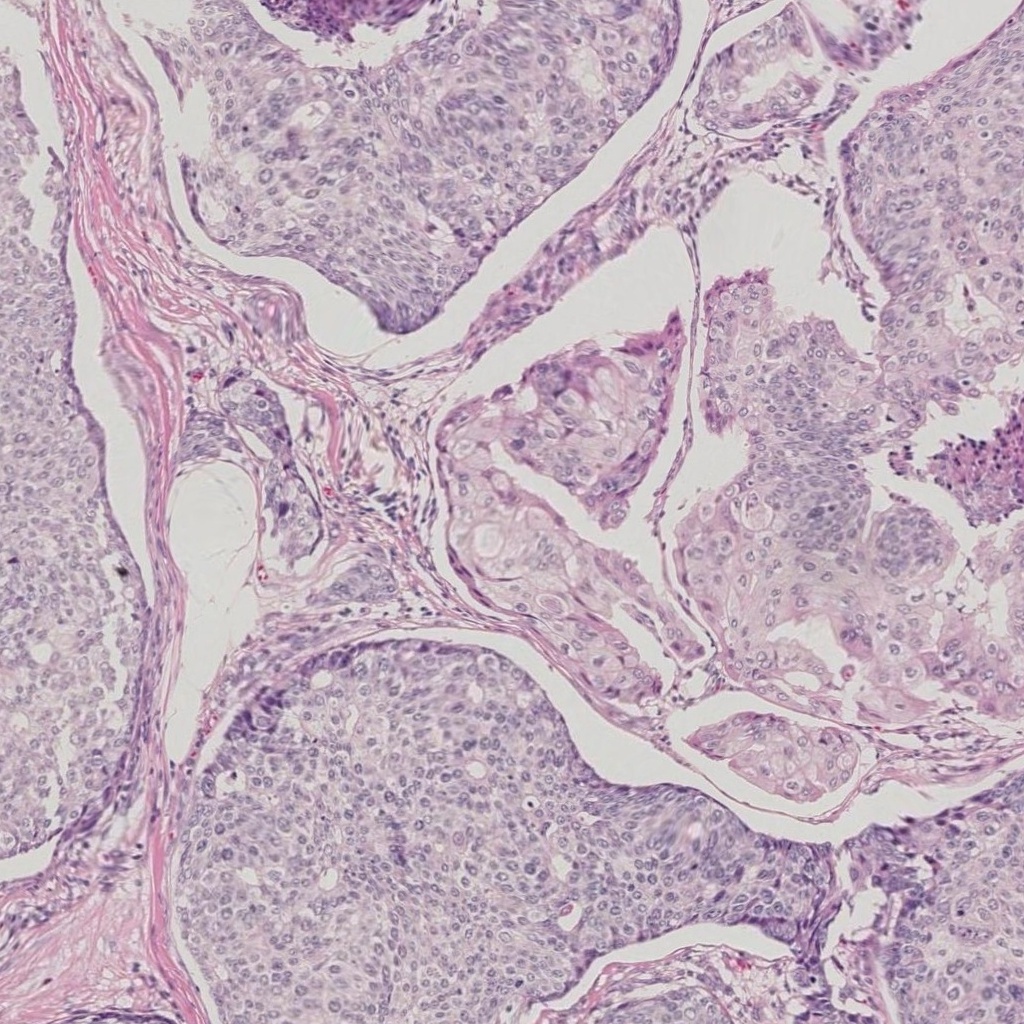} &
\includegraphics[width=1.5cm]{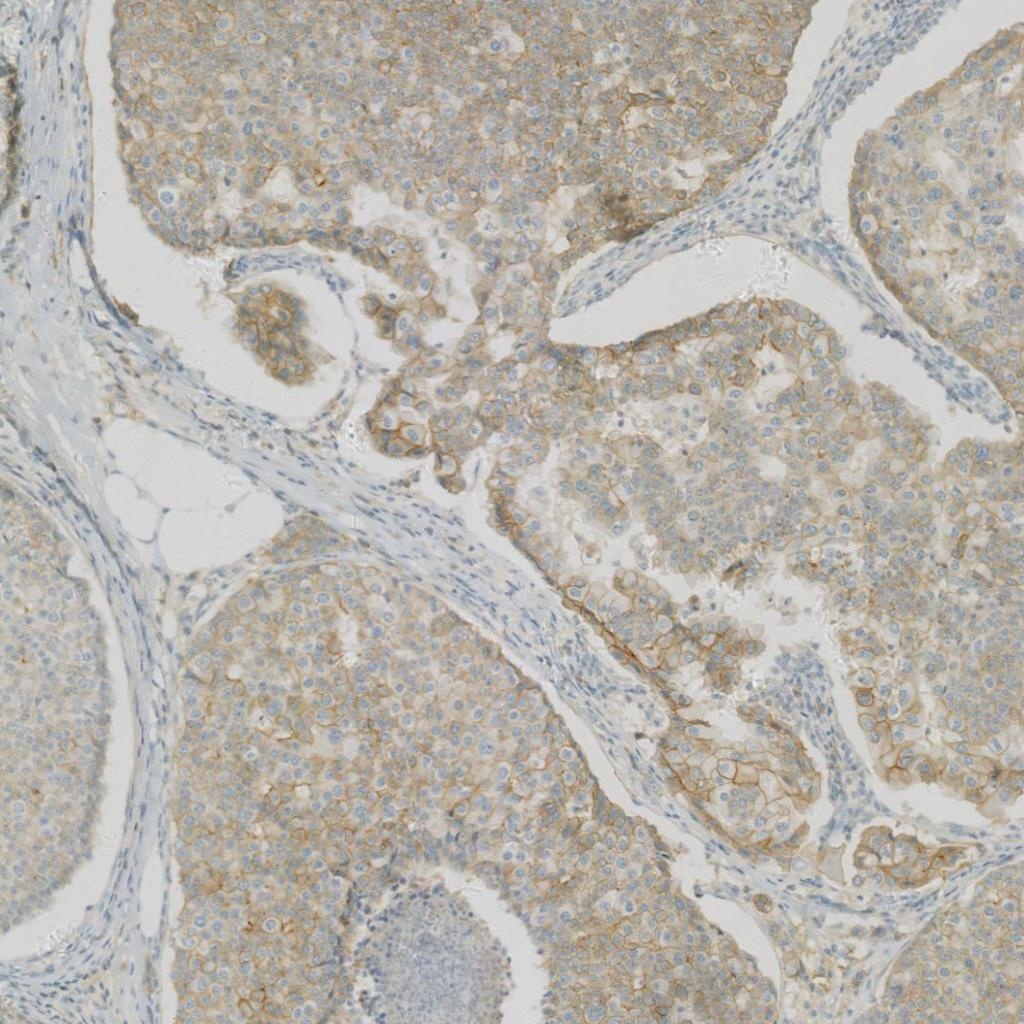} &
\includegraphics[width=1.5cm]{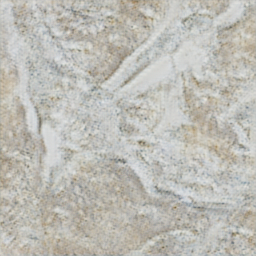} &
\includegraphics[width=1.5cm]{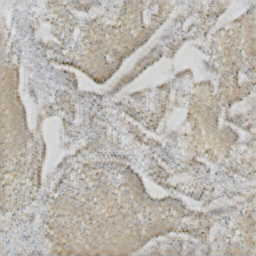} &
\includegraphics[width=1.5cm]{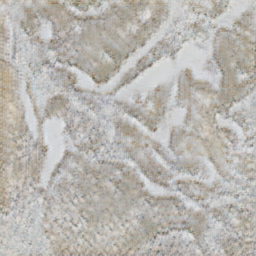} \\
\includegraphics[width=1.5cm]{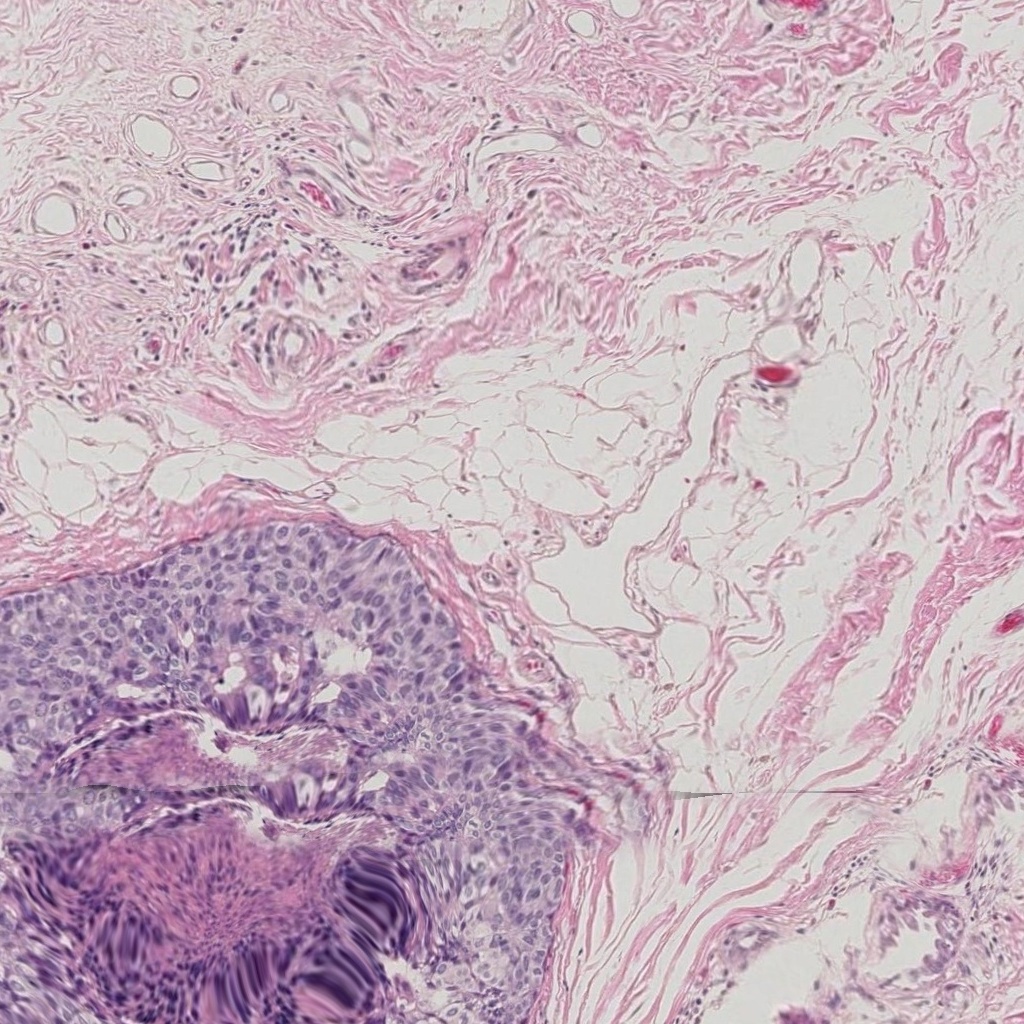} &
\includegraphics[width=1.5cm]{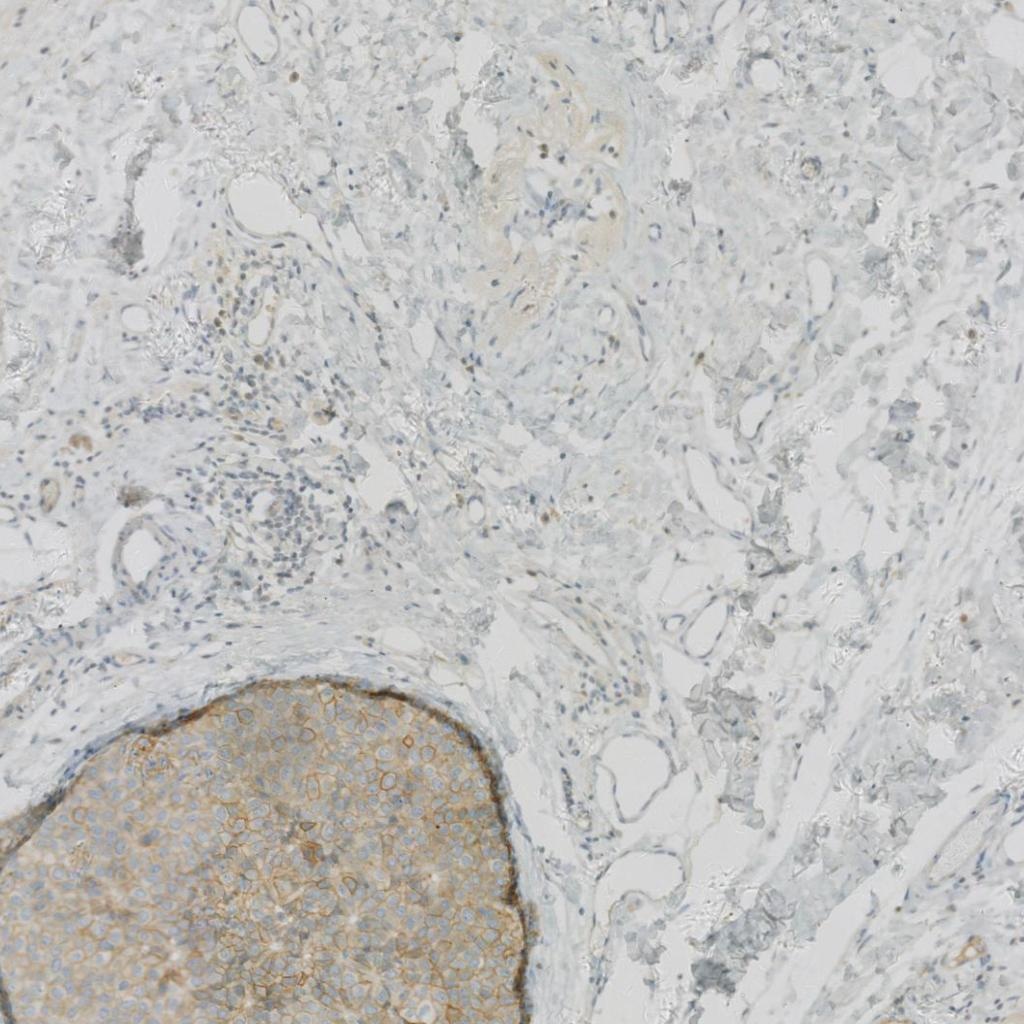} &
\includegraphics[width=1.5cm]{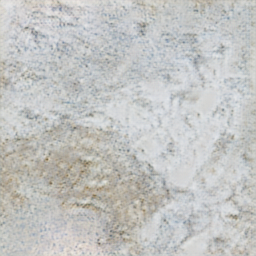} &
\includegraphics[width=1.5cm]{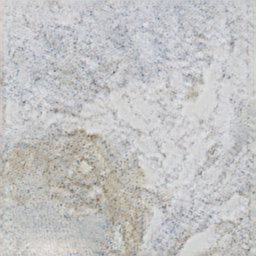} &
\includegraphics[width=1.5cm]{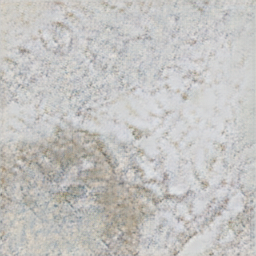} \\
\end{tabular}

\vspace{2mm}
{\small\textbf{(a)} Comparison of IHC 0 images generated using different methods.}

\vspace{6mm}

\begin{tabular}{>{\centering\arraybackslash}m{1.3cm}
                >{\centering\arraybackslash}m{1.3cm}
                >{\centering\arraybackslash}m{1.3cm}
                >{\centering\arraybackslash}m{1.3cm}
                >{\centering\arraybackslash}m{1.3cm}}
\small H\&E & \small Real IHC & \small pix2pix & \small pyramid pix2pix & \small ours\\
\includegraphics[width=1.5cm]{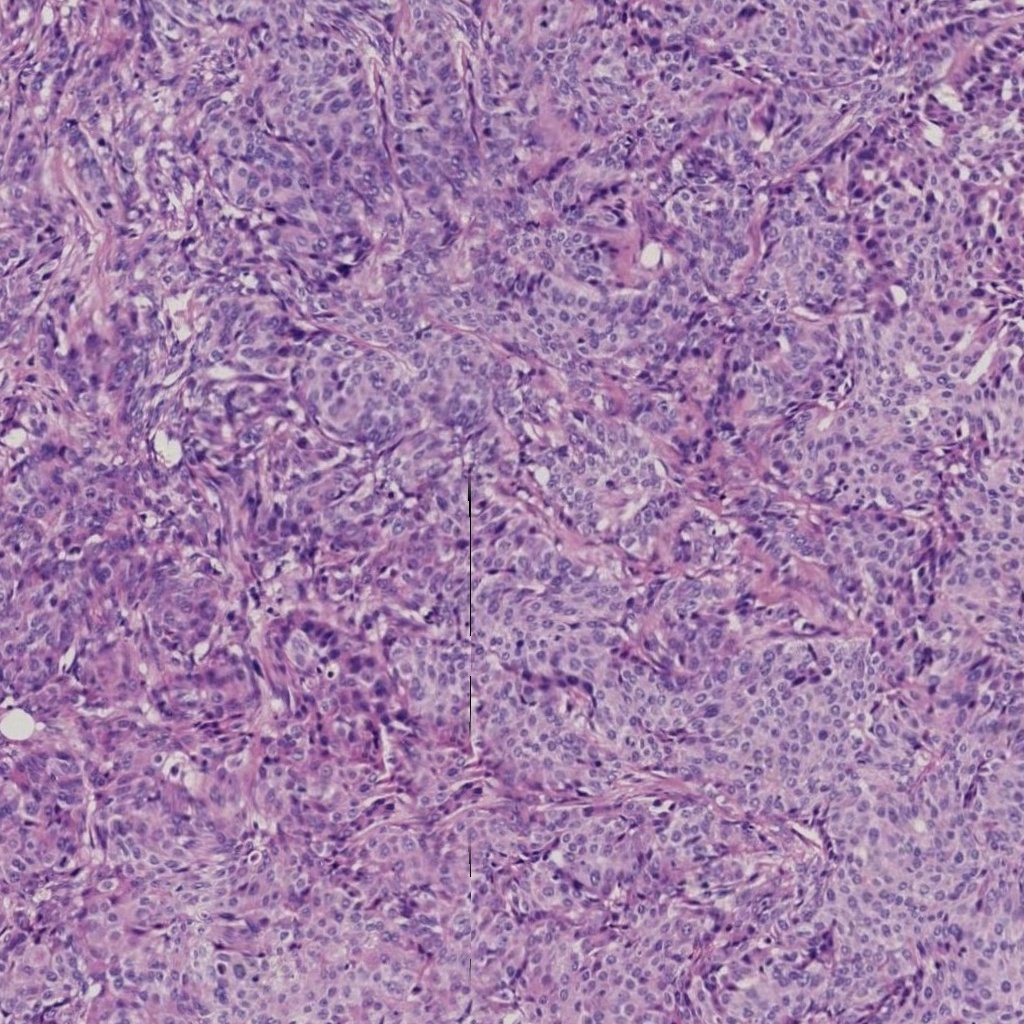} &
\includegraphics[width=1.5cm]{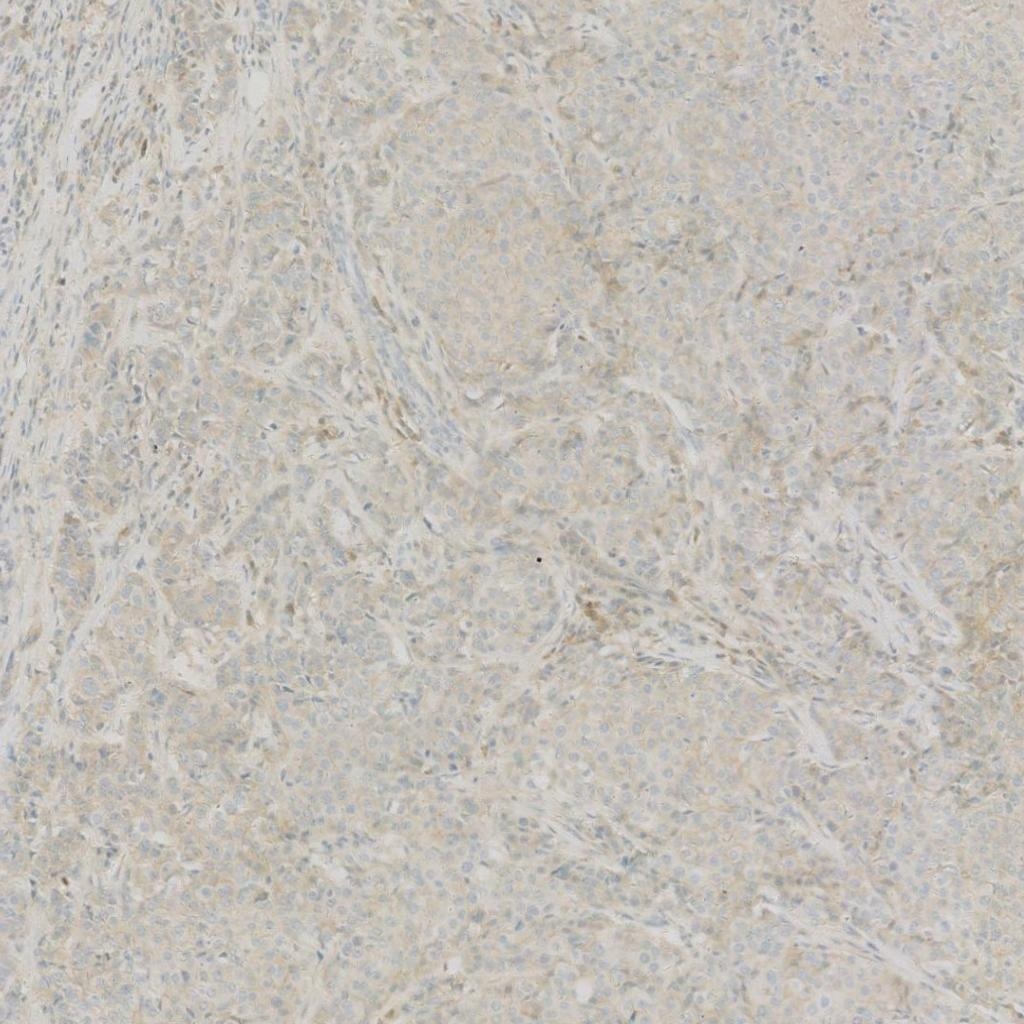} &
\includegraphics[width=1.5cm]{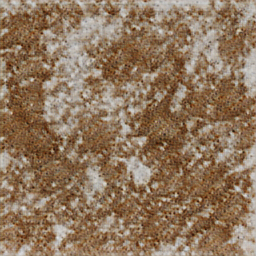} &
\includegraphics[width=1.5cm]{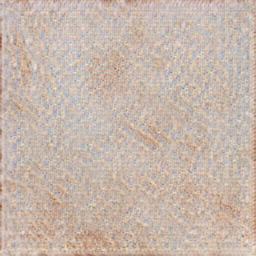} &
\includegraphics[width=1.5cm]{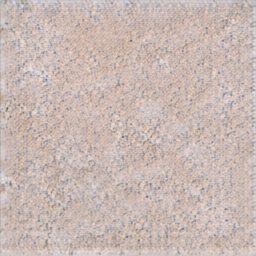} \\
\includegraphics[width=1.5cm]{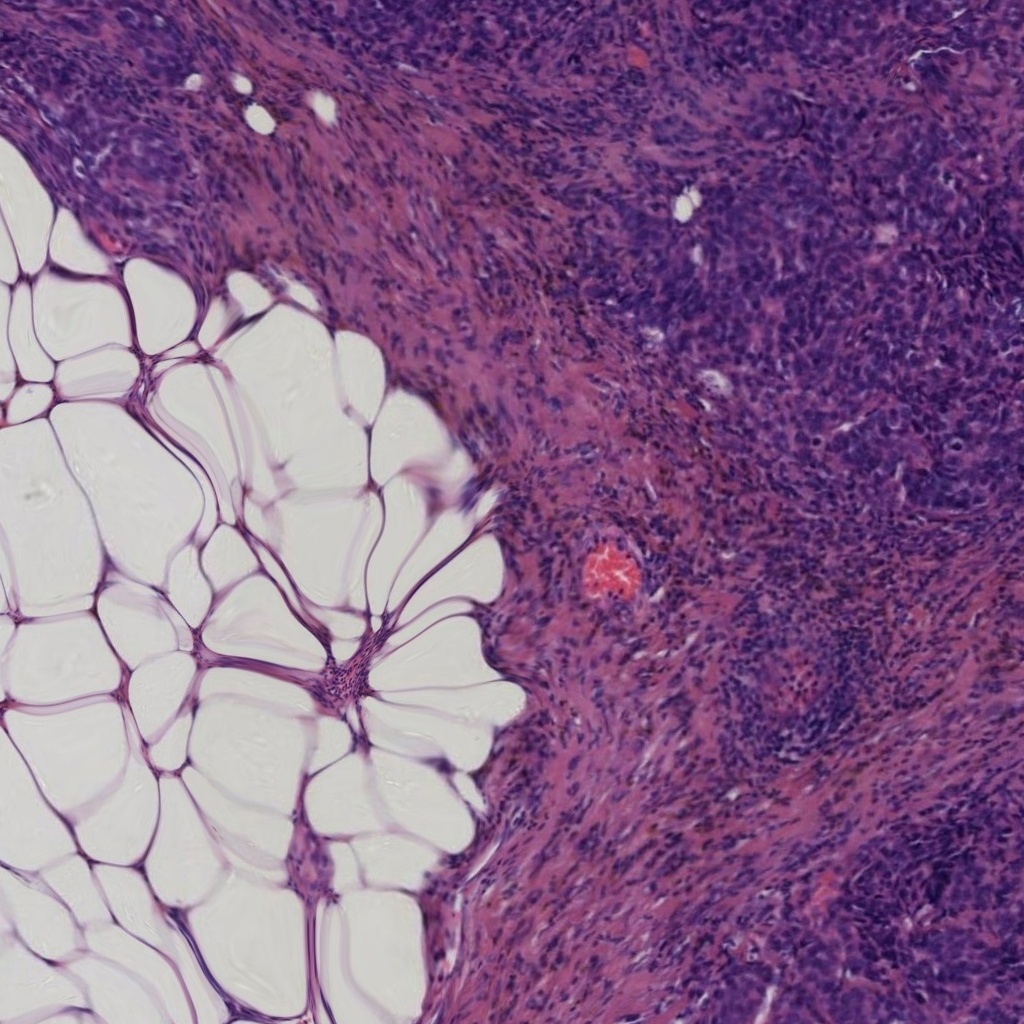} &
\includegraphics[width=1.5cm]{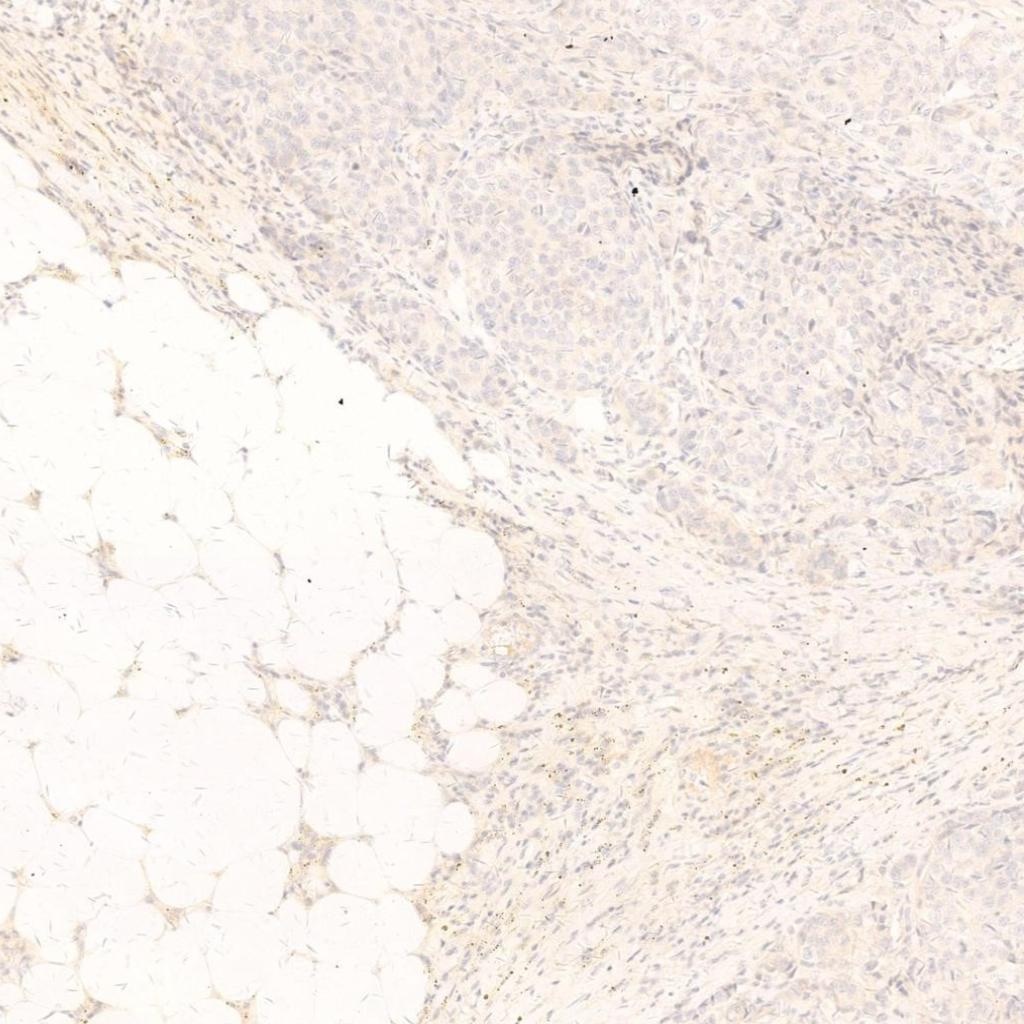} &
\includegraphics[width=1.5cm]{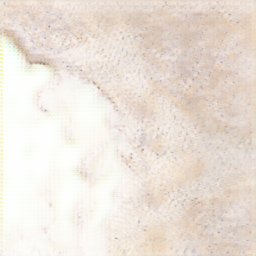} &
\includegraphics[width=1.5cm]{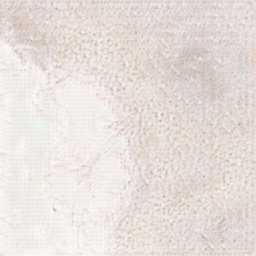} &
\includegraphics[width=1.5cm]{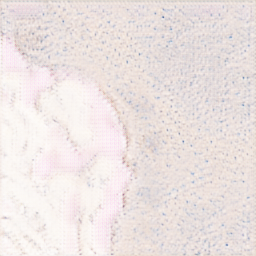} \\
\includegraphics[width=1.5cm]{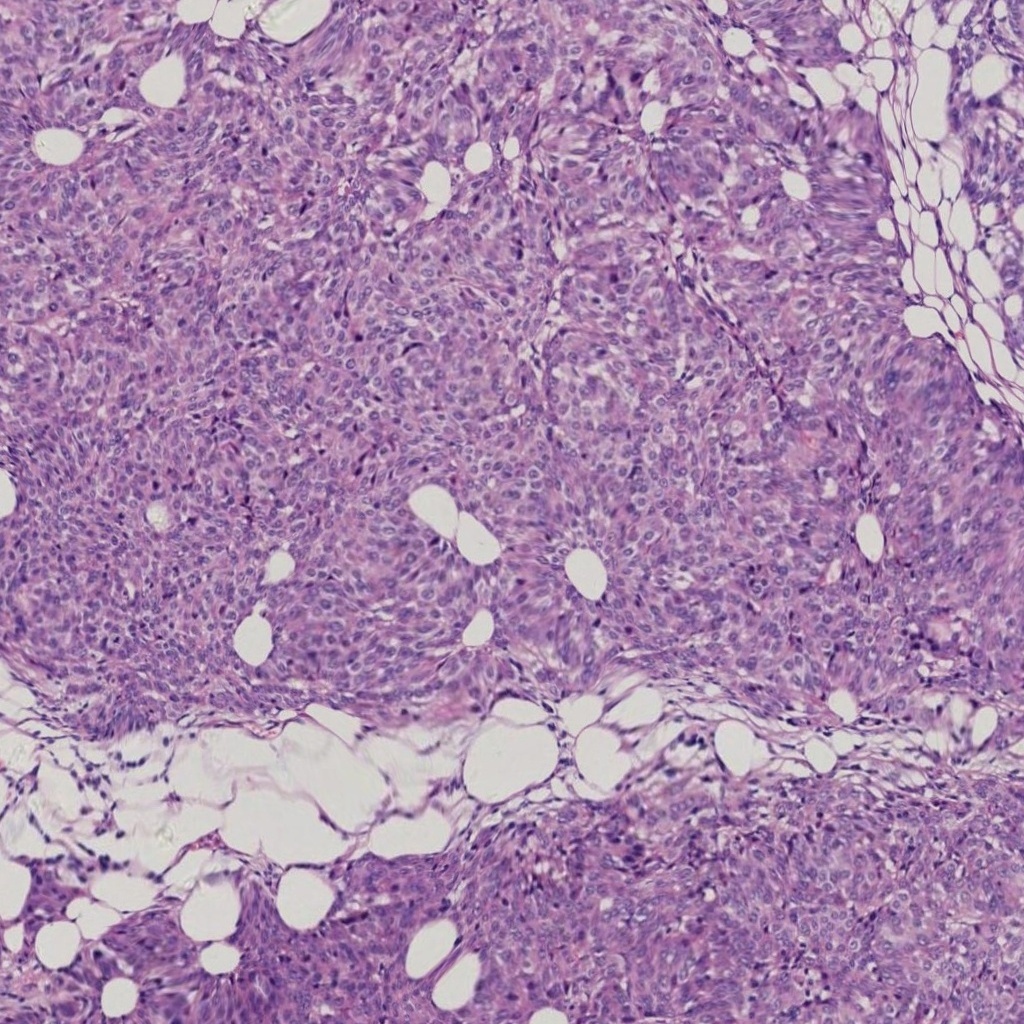} &
\includegraphics[width=1.5cm]{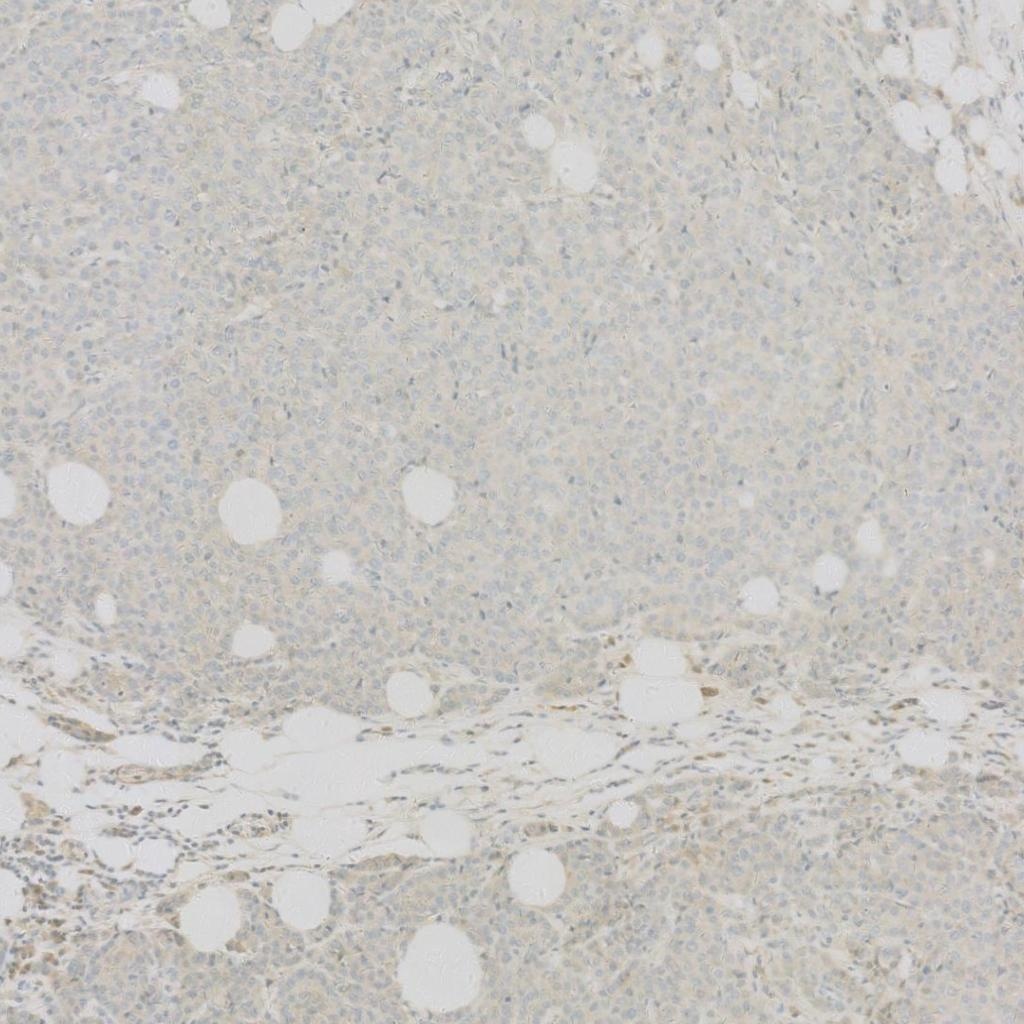} &
\includegraphics[width=1.5cm]{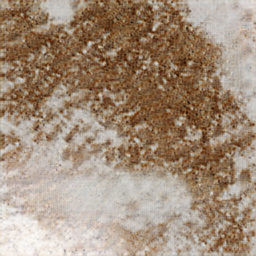} &
\includegraphics[width=1.5cm]{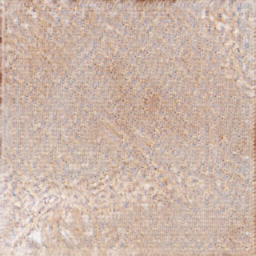} &
\includegraphics[width=1.5cm]{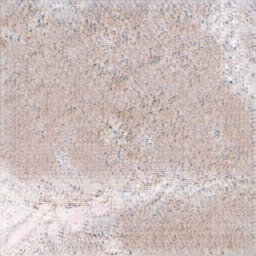} \\
\end{tabular}

\vspace{2mm}
{\small\textbf{(b)} Comparison of IHC 1+ images generated using different methods.}

\end{minipage}
\hfill
\begin{minipage}[t]{0.48\textwidth}
\centering

\begin{tabular}{>{\centering\arraybackslash}m{1.3cm}
                >{\centering\arraybackslash}m{1.3cm}
                >{\centering\arraybackslash}m{1.3cm}
                >{\centering\arraybackslash}m{1.3cm}
                >{\centering\arraybackslash}m{1.3cm}}
\small H\&E & \small Real IHC & \small pix2pix & \small pyramid pix2pix & \small ours\\
\includegraphics[width=1.5cm]{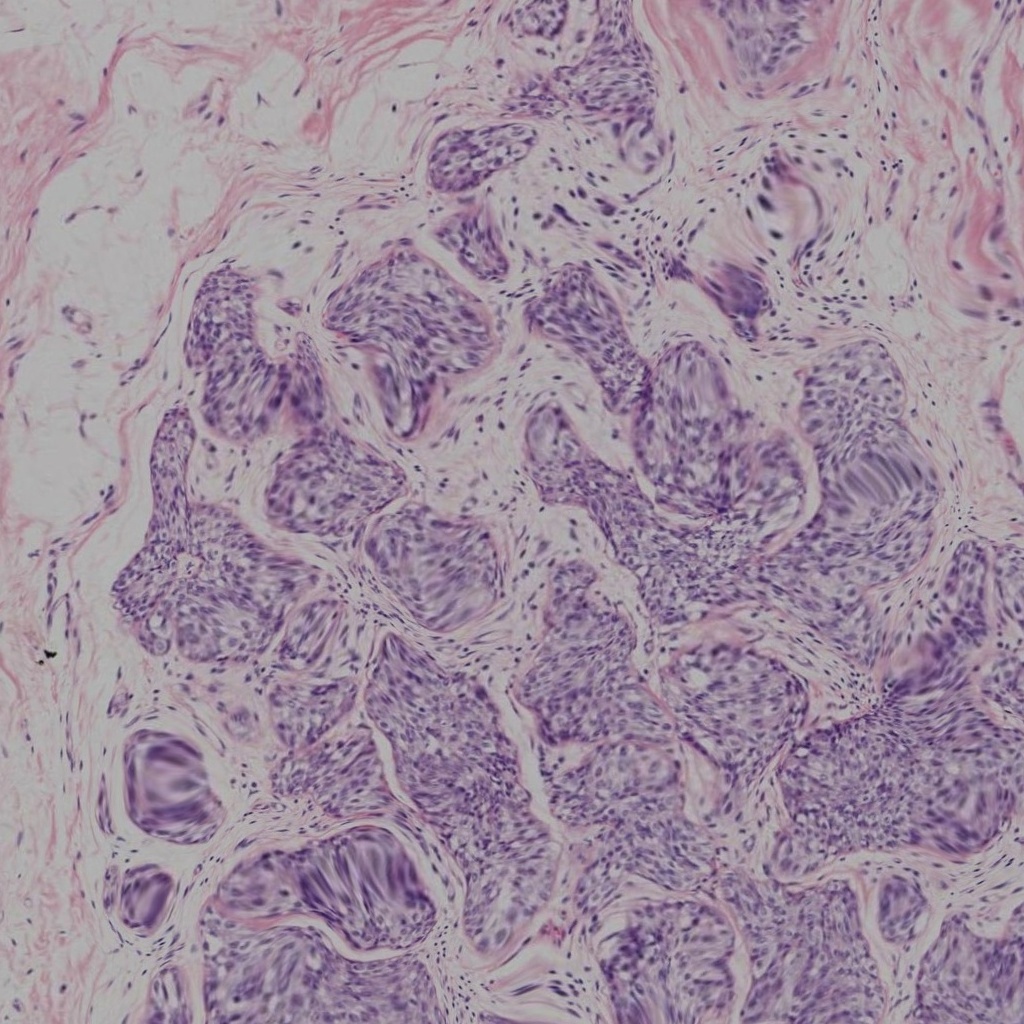} &
\includegraphics[width=1.5cm]{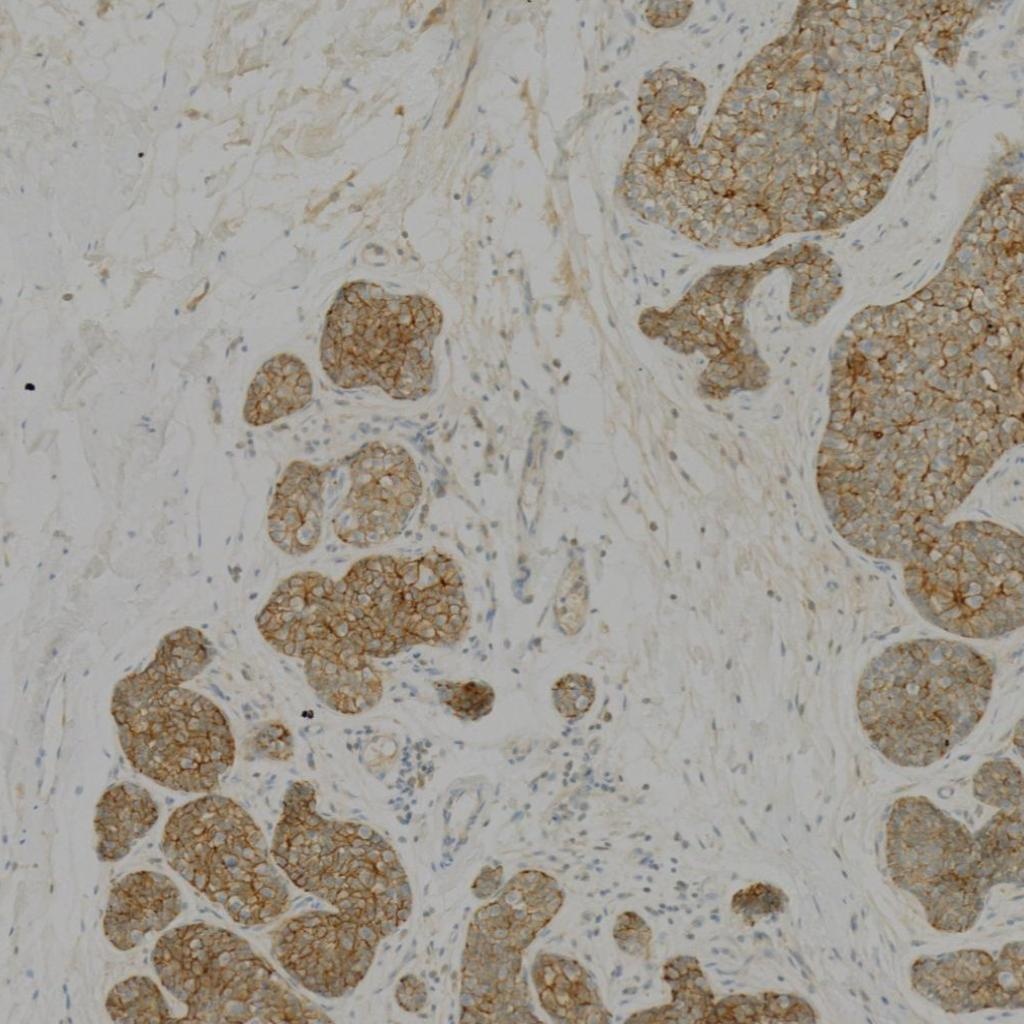} &
\includegraphics[width=1.5cm]{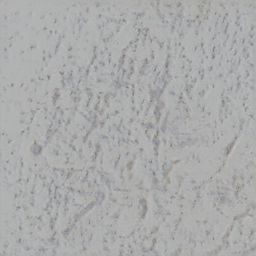} &
\includegraphics[width=1.5cm]{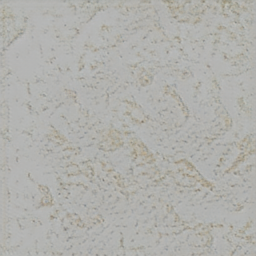} &
\includegraphics[width=1.5cm]{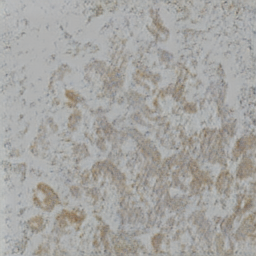} \\
\includegraphics[width=1.5cm]{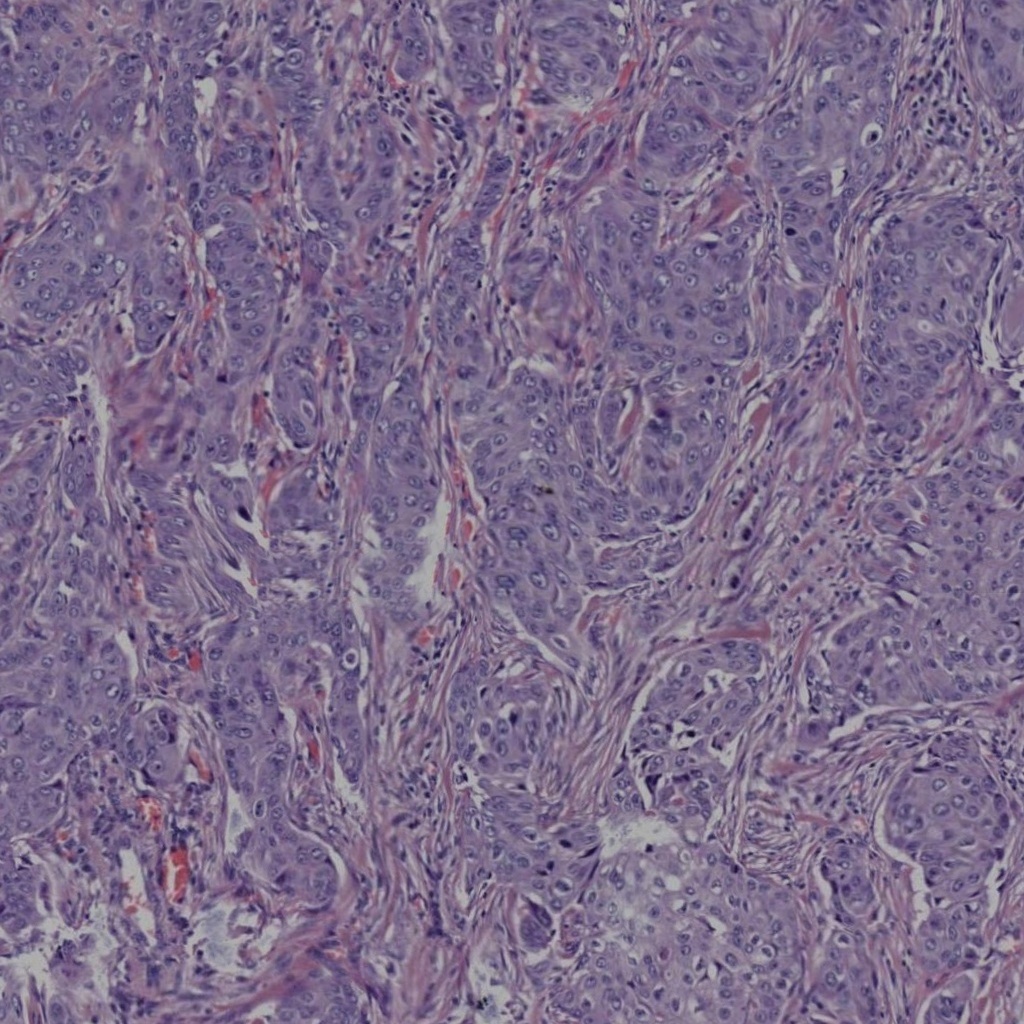} &
\includegraphics[width=1.5cm]{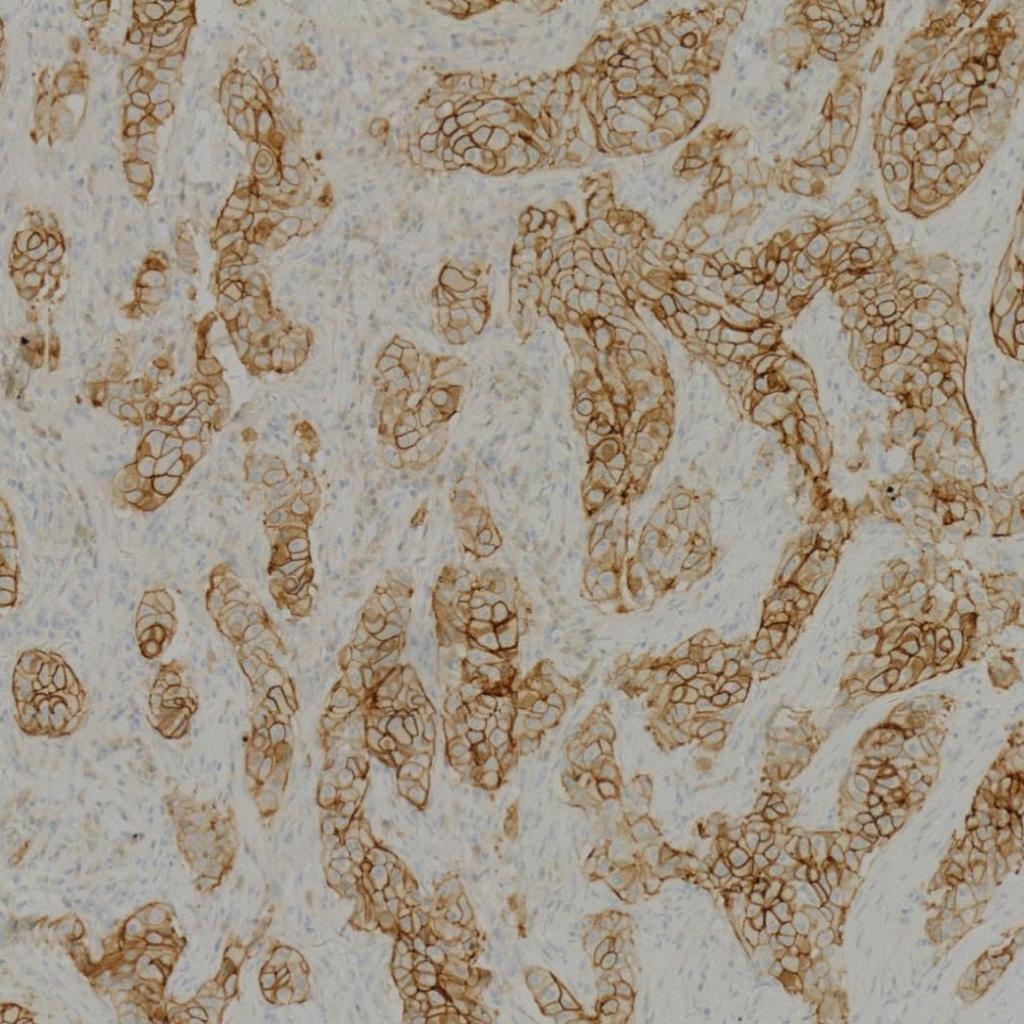} &
\includegraphics[width=1.5cm]{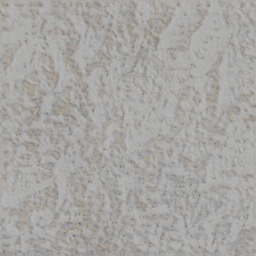} &
\includegraphics[width=1.5cm]{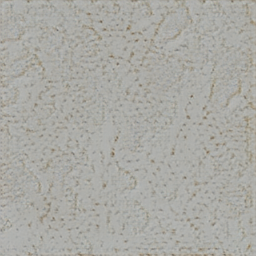} &
\includegraphics[width=1.5cm]{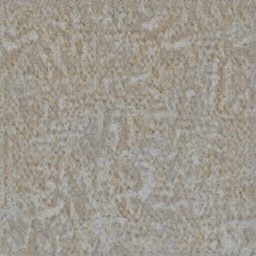} \\
\includegraphics[width=1.5cm]{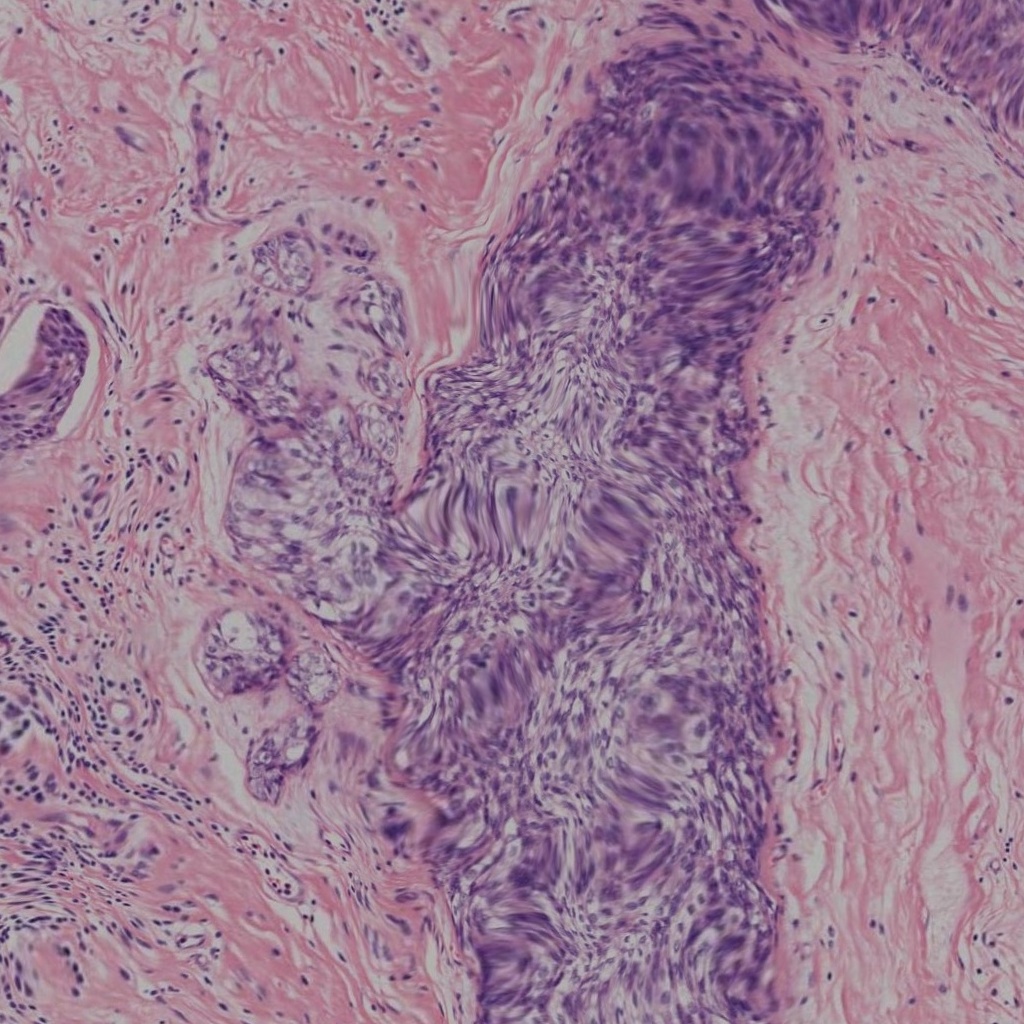} &
\includegraphics[width=1.5cm]{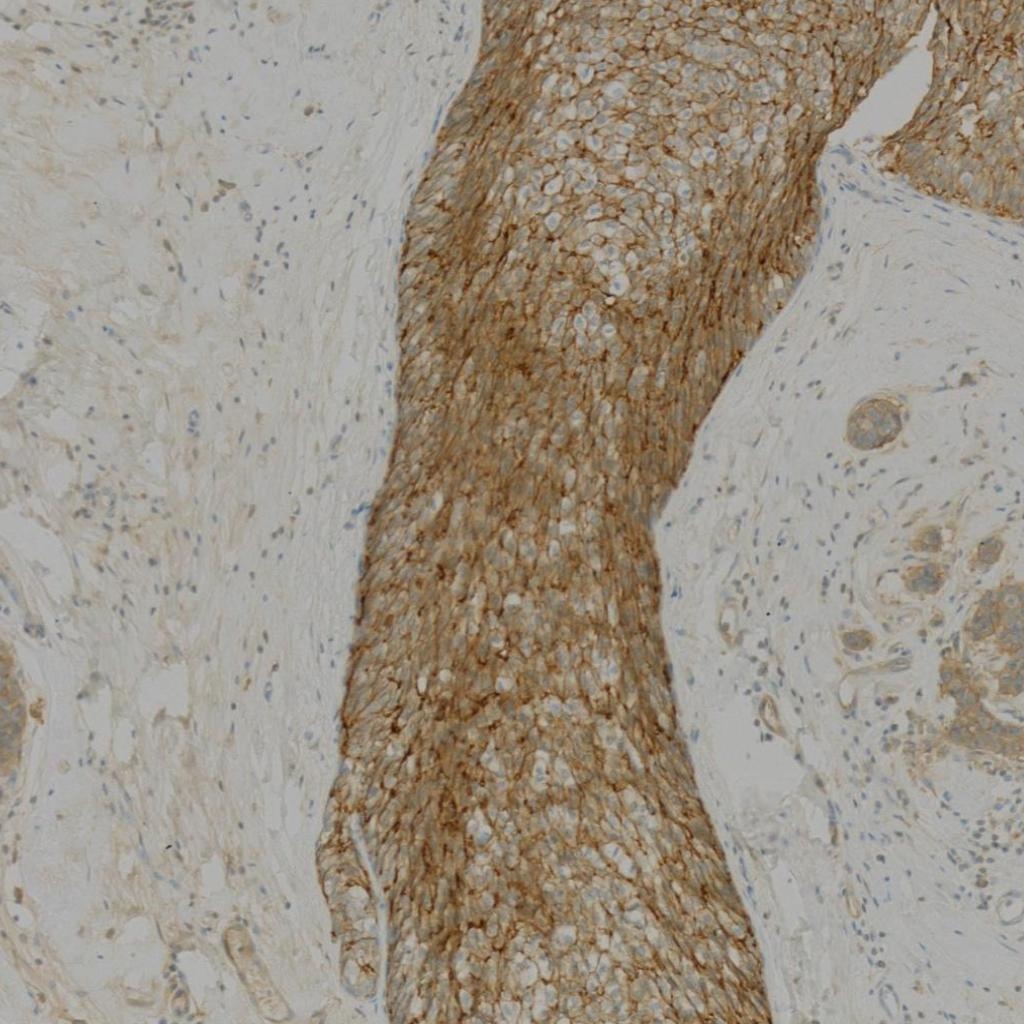} &
\includegraphics[width=1.5cm]{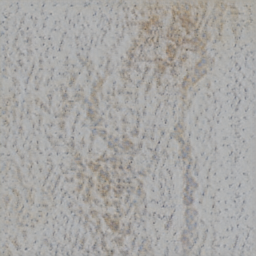} &
\includegraphics[width=1.5cm]{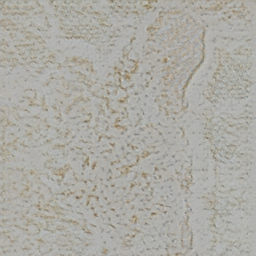} &
\includegraphics[width=1.5cm]{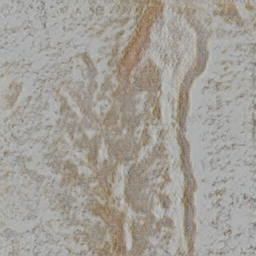} \\
\end{tabular}

\vspace{2mm}
{\small\textbf{(c)} Comparison of IHC 2+ images generated using different methods.}

\vspace{6mm}

\begin{tabular}{>{\centering\arraybackslash}m{1.3cm}
                >{\centering\arraybackslash}m{1.3cm}
                >{\centering\arraybackslash}m{1.3cm}
                >{\centering\arraybackslash}m{1.3cm}
                >{\centering\arraybackslash}m{1.3cm}}
\small H\&E & \small Real IHC & \small pix2pix & \small pyramid pix2pix & \small ours\\
\includegraphics[width=1.5cm]{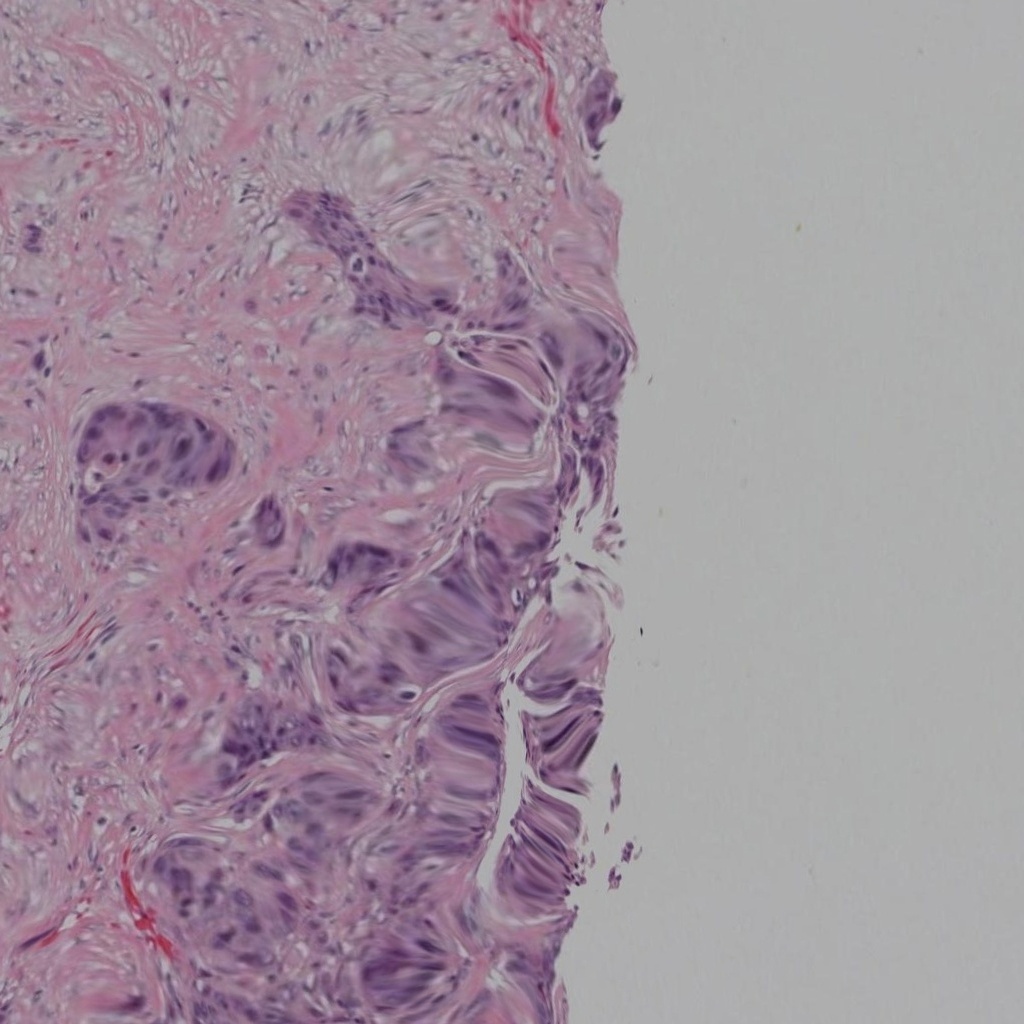} &
\includegraphics[width=1.5cm]{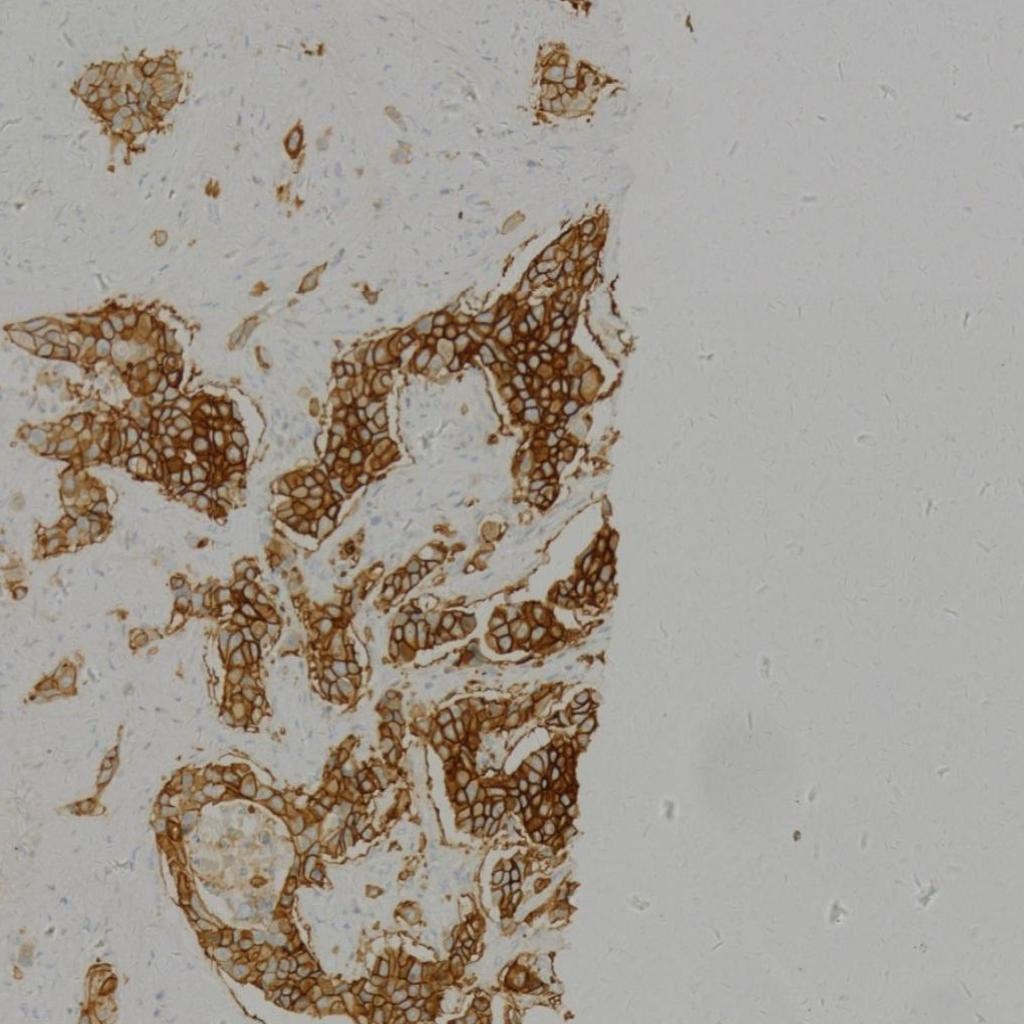} &
\includegraphics[width=1.5cm]{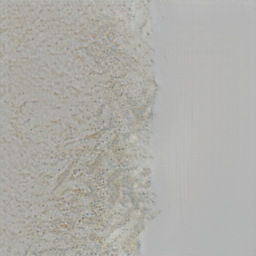} &
\includegraphics[width=1.5cm]{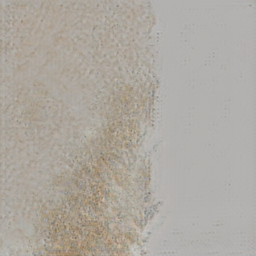} &
\includegraphics[width=1.5cm]{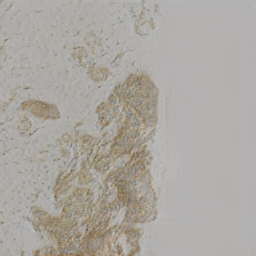} \\
\includegraphics[width=1.5cm]{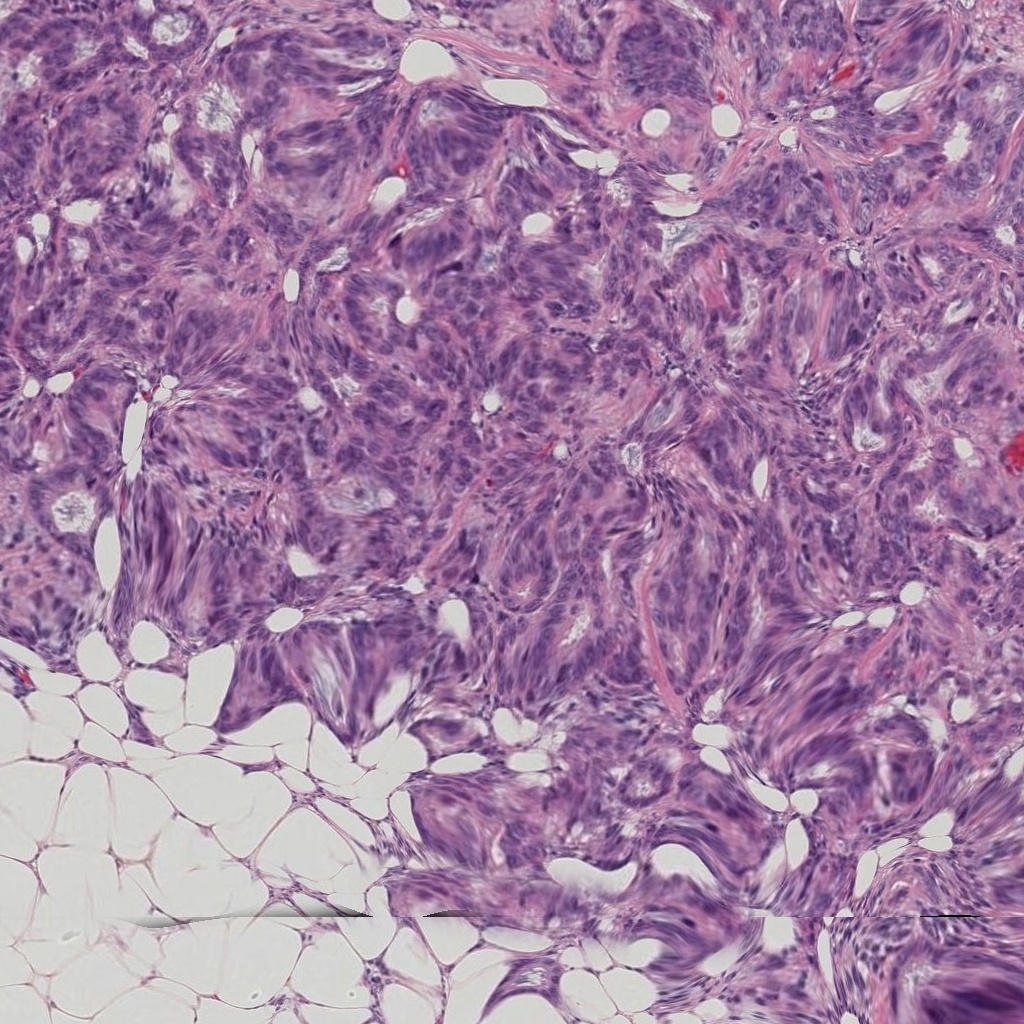} &
\includegraphics[width=1.5cm]{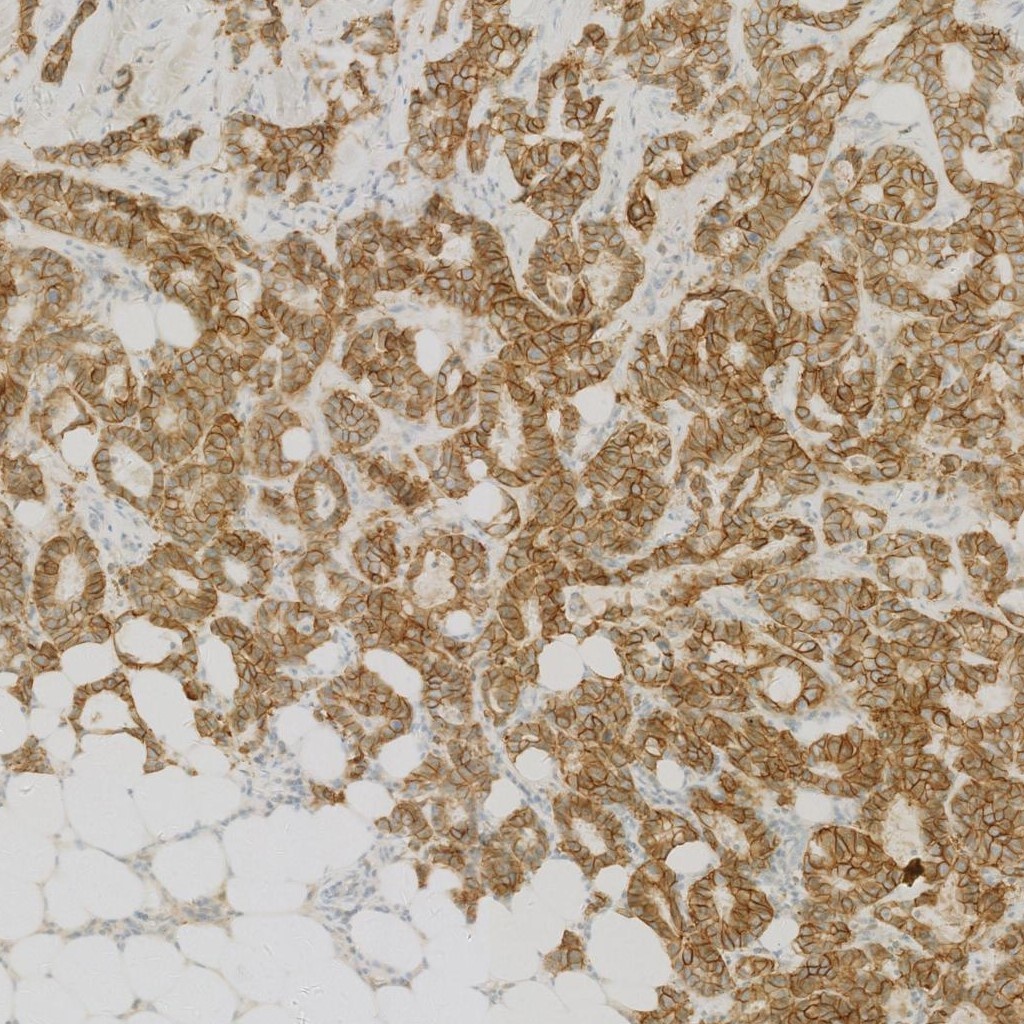} &
\includegraphics[width=1.5cm]{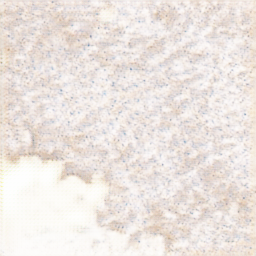} &
\includegraphics[width=1.5cm]{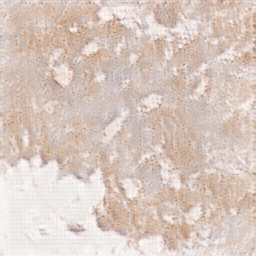} &
\includegraphics[width=1.5cm]{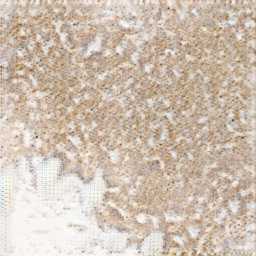} \\
\includegraphics[width=1.5cm]{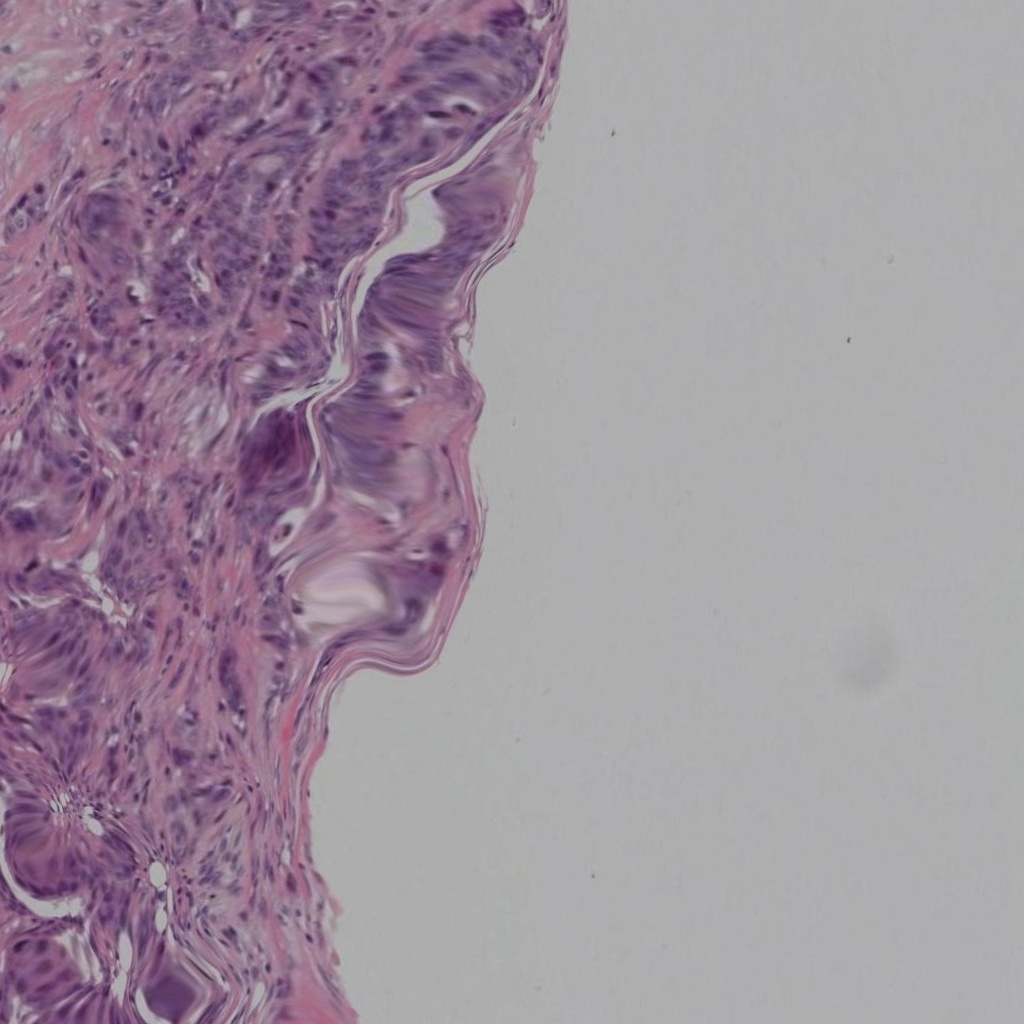} &
\includegraphics[width=1.5cm]{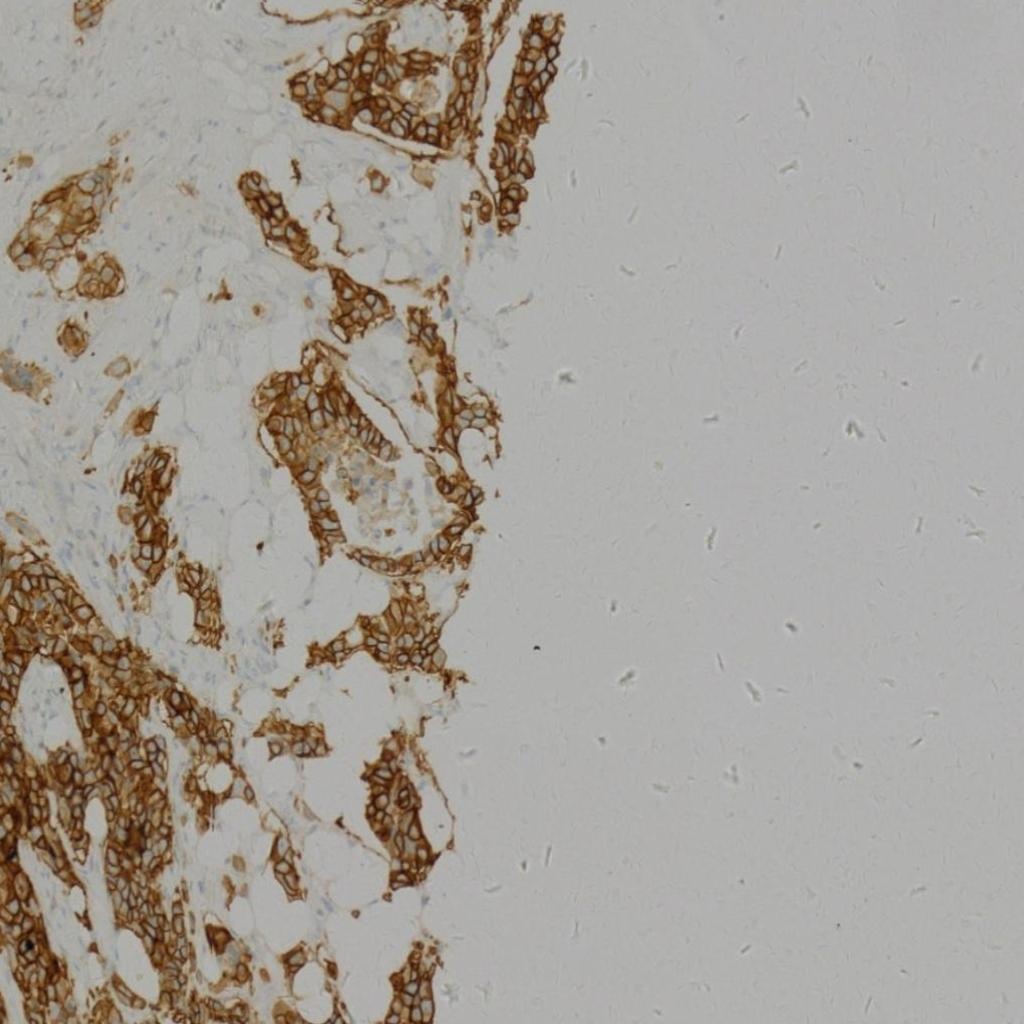} &
\includegraphics[width=1.5cm]{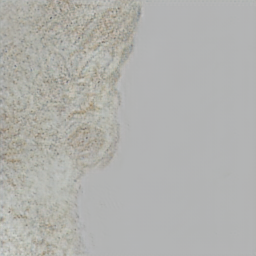} &
\includegraphics[width=1.5cm]{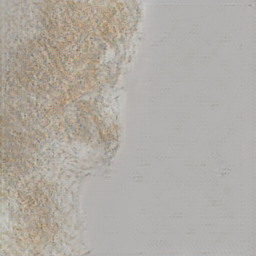} &
\includegraphics[width=1.5cm]{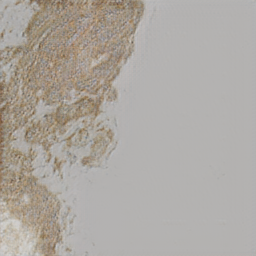} \\
\end{tabular}

\vspace{2mm}
{\small\textbf{(d)} Comparison of IHC 3+ images generated using different methods.}

\end{minipage}

\caption{Comparison of different types of IHC images generated using different methods. The first column shows three randomly chosen input H\&E images, the second column shows the corresponding real IHC images. The third, fourth and fifth columns represent IHC images generated using pix2pix, pyramid pix2pix and our method respectively. The images generated by our method are closer to the real IHC images.}
\label{fig:comparison2}
\end{figure*}

Additionally, Figure \ref{fig:comparison2} presents a qualitative comparison of IHC images generated for various HER2 expression levels using different models. The images generated by our approach more closely resemble the ground truth IHC images across all HER2 classes, with a particularly noticeable improvement for IHC 3+ cases. This substantiates our model’s ability to translate H\&E images into realistic IHC representations, reinforcing its utility in medical image analysis and diagnostic workflows.

\subsubsection{Training Dynamics and Convergence Analysis}

\begin{figure}[htbp]
    \centering
    
    \begin{minipage}{0.48\linewidth}
        \centering
        \includegraphics[width=\linewidth]{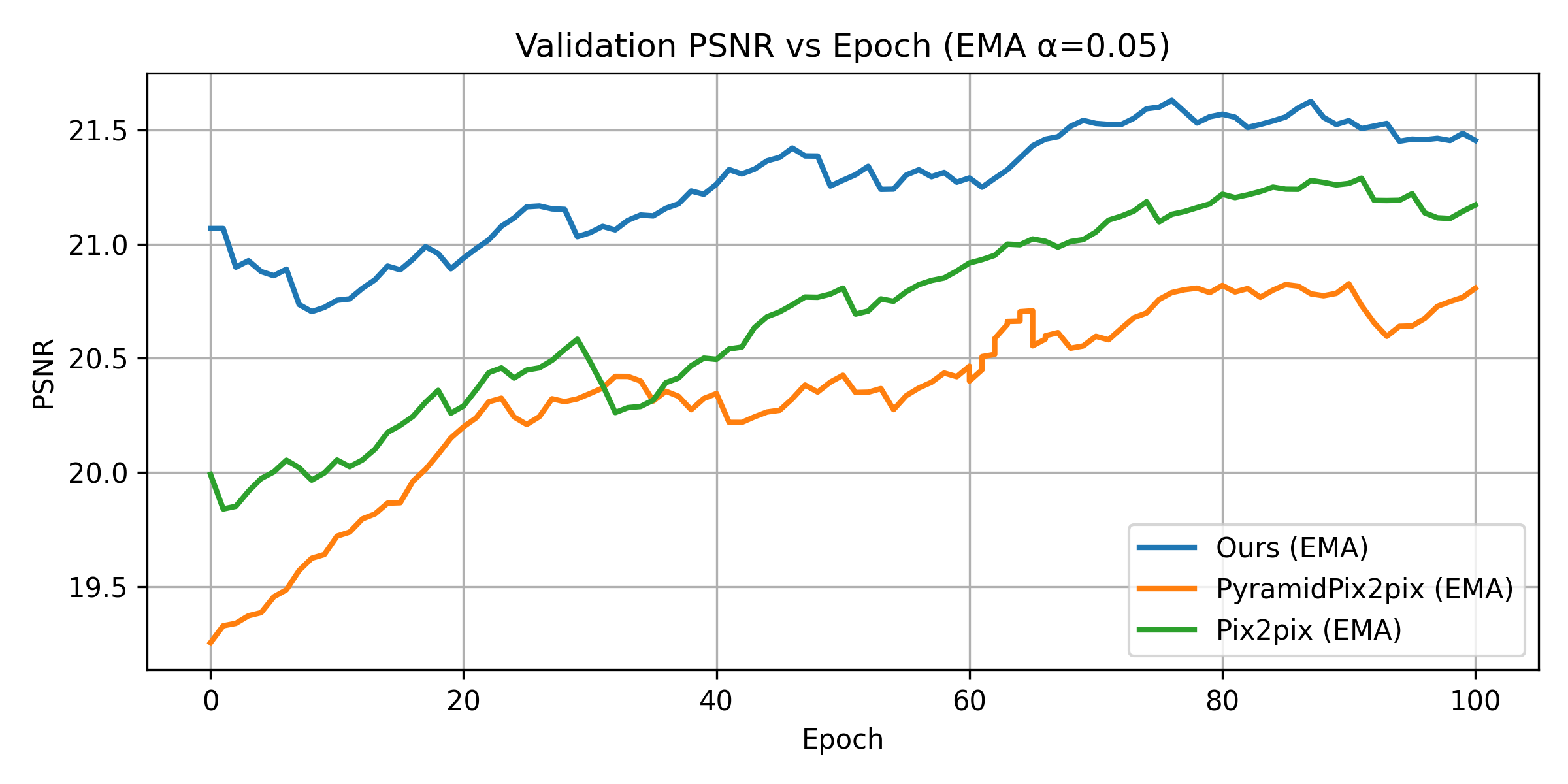}
        \vspace{2mm}
        \small (a) Validation PSNR
    \end{minipage}
    \hfill
    \begin{minipage}{0.48\linewidth}
        \centering
        \includegraphics[width=\linewidth]{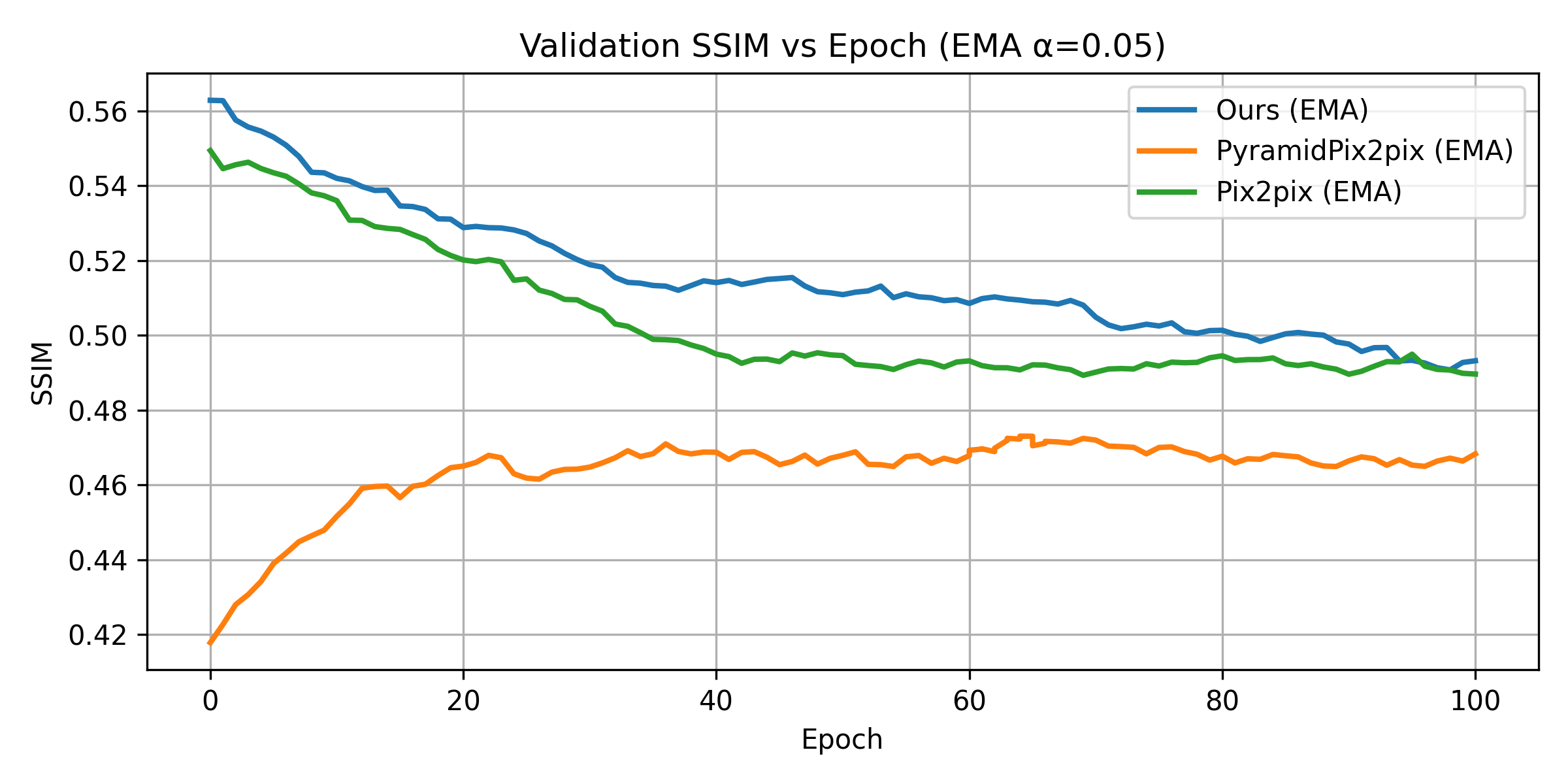}
        \vspace{2mm}
        \small (b) Validation SSIM
    \end{minipage}
    
    \caption{Validation performance comparison of the proposed method, PyramidPix2pix, and Pix2pix over 100 training epochs. Exponential moving average (EMA, $\alpha = 0.05$) is applied to highlight convergence trends. The proposed method consistently achieves superior PSNR and SSIM compared to the baseline models.}
    
    \label{fig:val_metrics}
\end{figure}

We compared the proposed variance-regularized method with Pix2pix and PyramidPix2pix on the validation set using PSNR and SSIM over 100 training epochs. As shown in Figure \ref{fig:val_metrics}, the proposed method consistently achieves higher PSNR values than both baselines throughout training. The improvement remains stable across epochs, with a clear margin over Pix2pix and PyramidPix2pix. In addition, the PSNR curve of our method exhibits smoother and more stable convergence behavior.
The higher PSNR indicates improved pixel-level reconstruction accuracy. By enforcing variance consistency between real and generated images at each spatial location, the model better preserves intensity dispersion and avoids excessive smoothing.
Figure X(b) presents the SSIM comparison. The proposed method maintains the highest SSIM values for most of the training process and achieves superior structural similarity at convergence.

A slight decrease in SSIM is observed during early epochs. This behavior is expected, as the variance regularization increases local intensity fluctuations while the generator adapts to match real-image variance statistics. After this adaptation phase, SSIM stabilizes and remains consistently higher than the baseline methods.

\begin{figure}[htbp]
\centering

\begin{minipage}{0.48\linewidth}
    \centering
    \includegraphics[width=\linewidth]{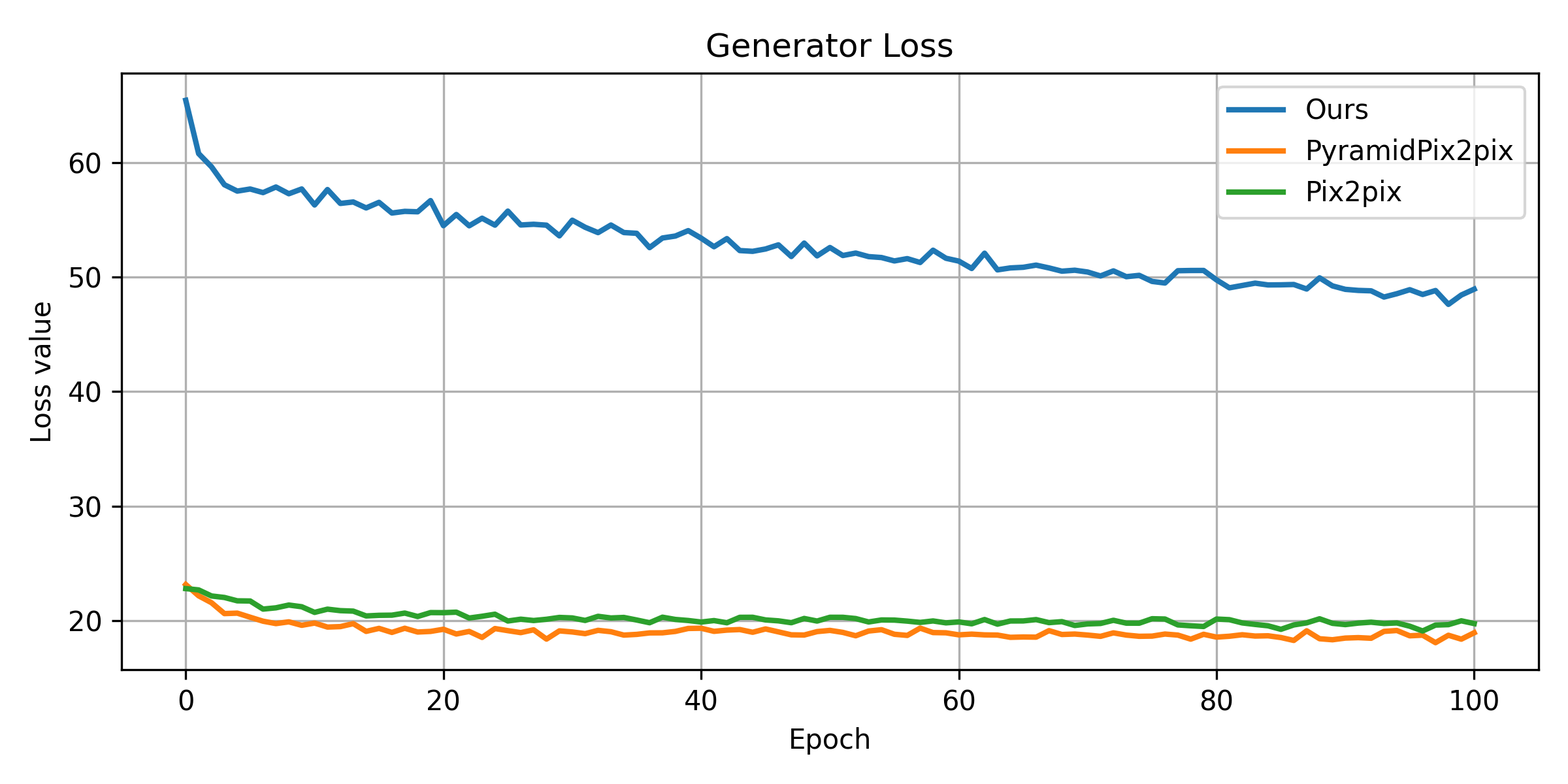}
    \vspace{1mm}
    \small (a) Generator Loss
\end{minipage}
\hfill
\begin{minipage}{0.48\linewidth}
    \centering
    \includegraphics[width=\linewidth]{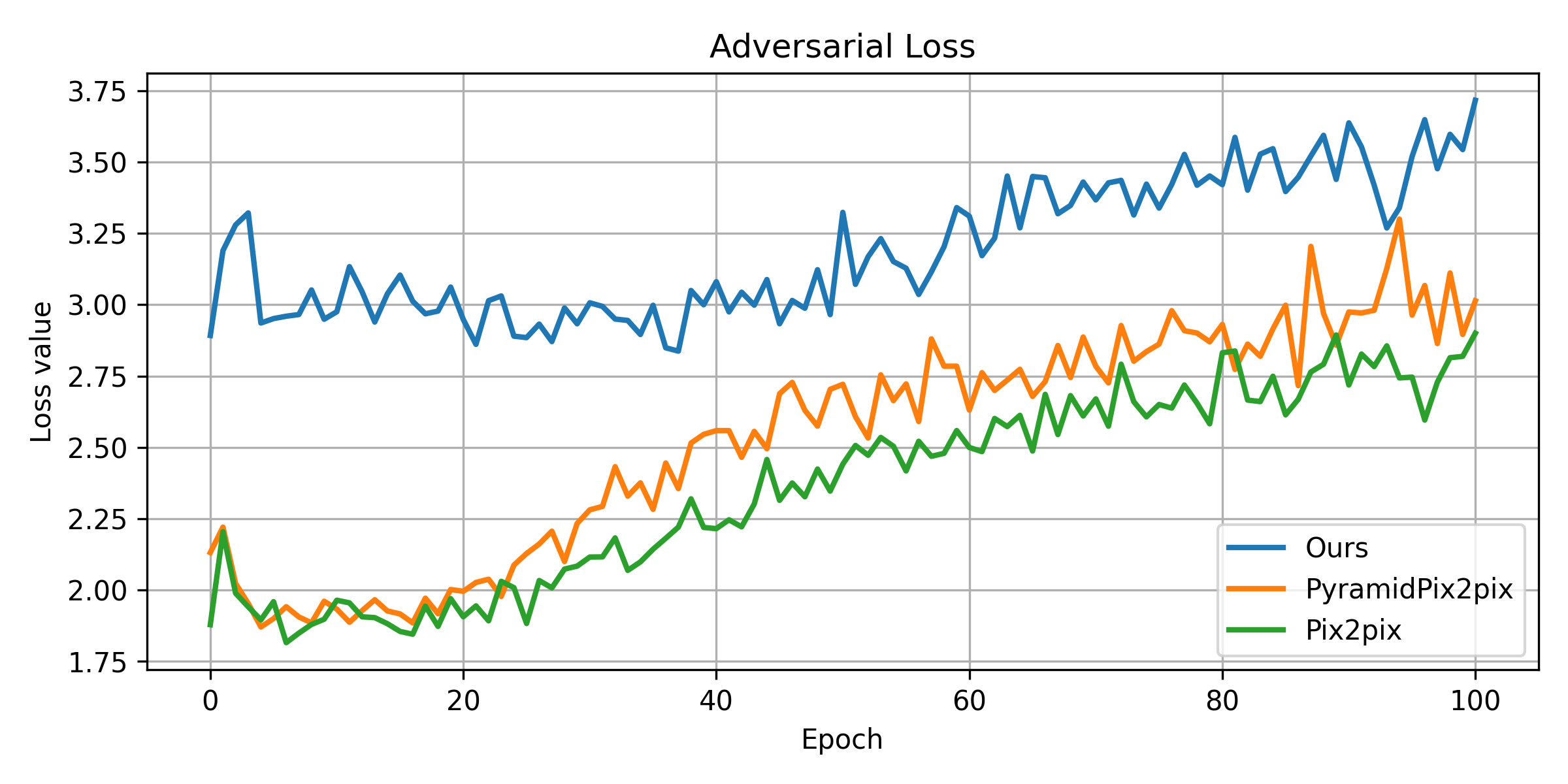}
    \vspace{1mm}
    \small (b) Adversarial Loss
\end{minipage}

\vspace{3mm}

\begin{minipage}{0.48\linewidth}
    \centering
    \includegraphics[width=\linewidth]{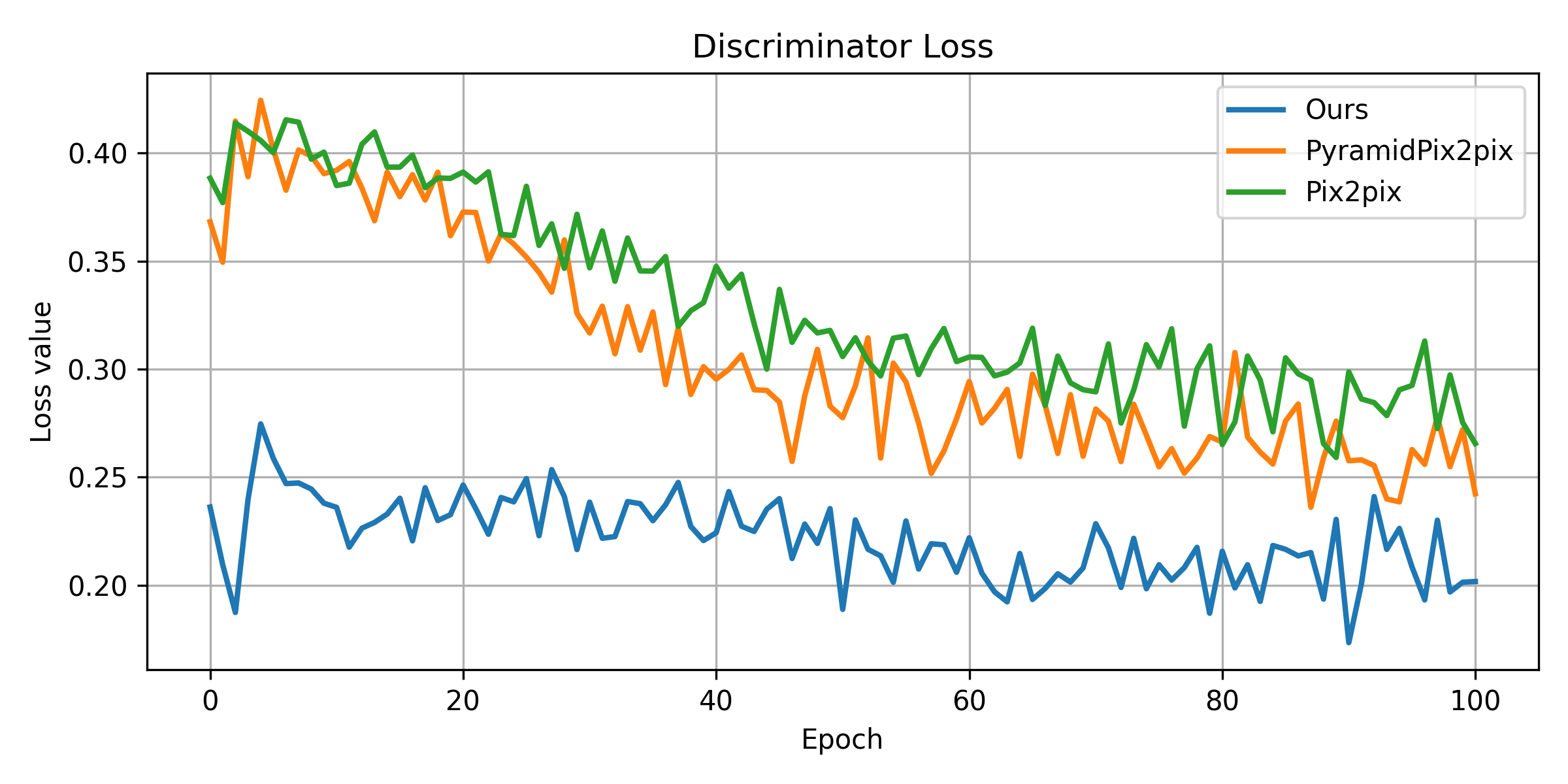}
    \vspace{1mm}
    \small (c) Discriminator Loss
\end{minipage}
\hfill
\begin{minipage}{0.48\linewidth}
    \centering
    \includegraphics[width=\linewidth]{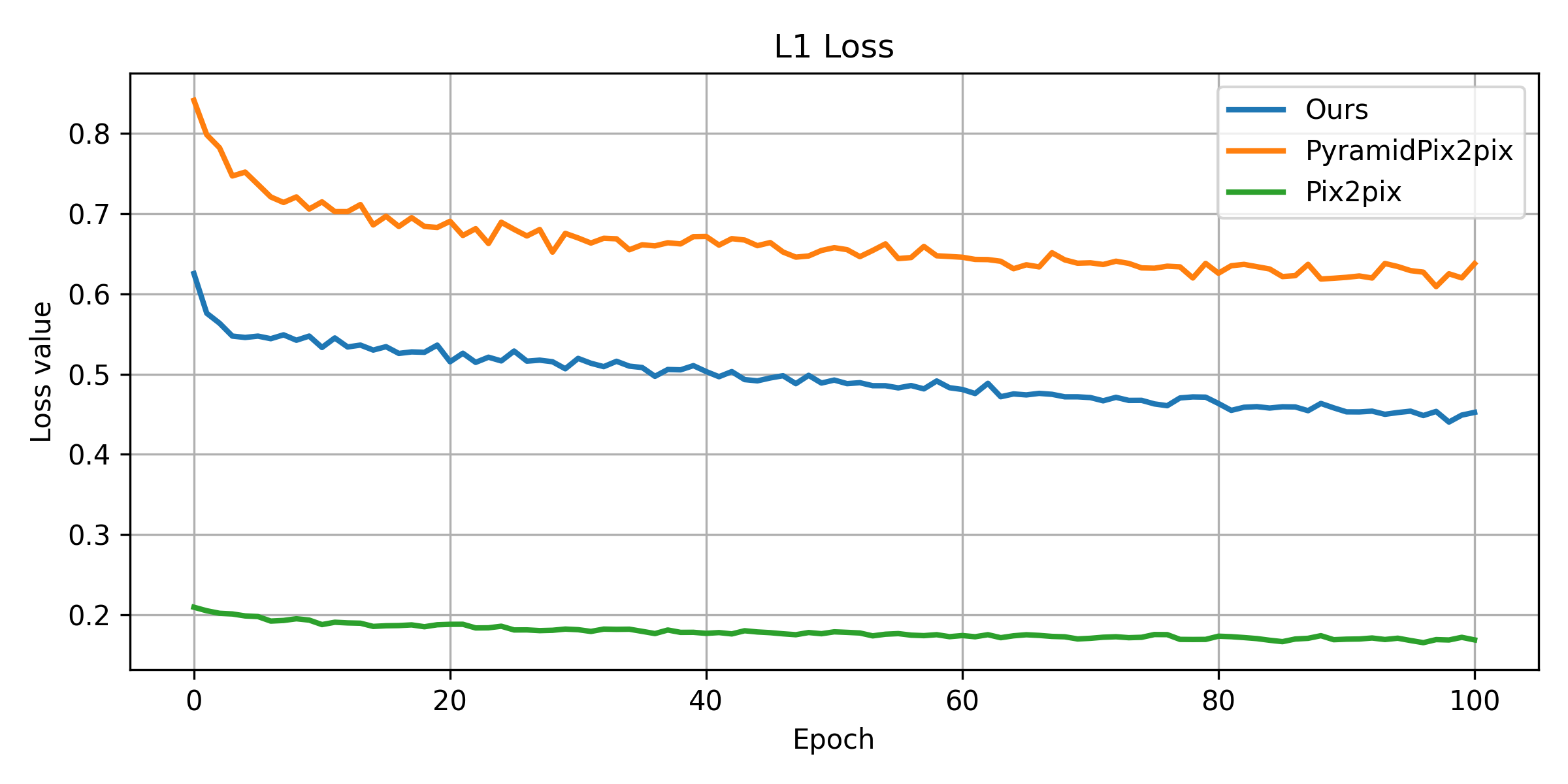}
    \vspace{1mm}
    \small (d) L1 Loss
\end{minipage}

\caption{Training loss comparison across the proposed method, Pix2pix, and PyramidPix2pix over 100 epochs. (a) Total generator loss, (b) adversarial loss, (c) discriminator loss, and (d) L1 reconstruction loss. The proposed method incorporates an additional variance-based regularization term, resulting in higher generator objective magnitude while maintaining stable discriminator dynamics and balanced adversarial training.}
\label{fig:loss_comparison}
\end{figure}

Figure \ref{fig:loss_comparison} compares different training losses for our method against pyramidPix2pix and pix2pix. The increased generator loss magnitude arises from the additional variance regularization term and does not indicate inferior adversarial performance. Instead, its smooth convergence demonstrates stable multi-objective optimization.The discriminator loss remains stable and bounded throughout training, indicating balanced adversarial optimization without discriminator saturation or collapse.The gradual increase in adversarial loss reflects a strengthening discriminator and a more challenging adversarial game, rather than degradation in generated image quality. This is supported by simultaneous improvements in PSNR and SSIM.Although Pix2pix achieves lower L1 loss, this primarily reflects stronger pixel-wise averaging. The proposed method achieves superior PSNR and SSIM despite a higher L1 term, indicating improved structural and statistical fidelity rather than over-smoothed reconstruction.

\begin{figure}[htbp]
    \centering
    \includegraphics[width=0.5\linewidth]{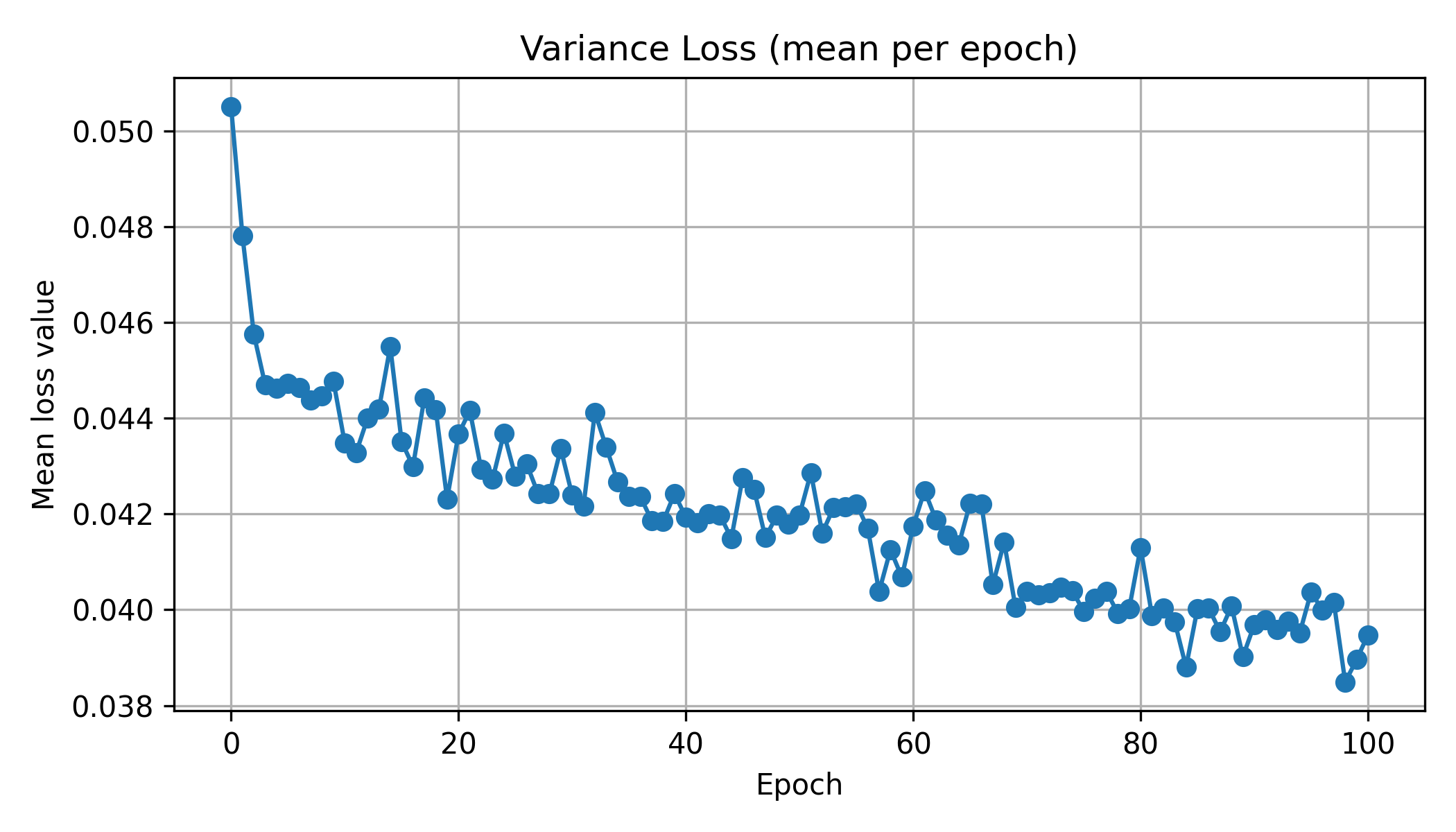}
    \caption{Convergence of the variance regularization loss over 100 training epochs, showing stable and consistent optimization.}
    \label{fig:var_loss}
\end{figure}

Figure \ref{fig:var_loss} shows the mean variance loss per epoch we proposed in our method. The steady downward trend and absence of oscillatory instability demonstrate that the variance-based regularization integrates smoothly into the adversarial framework and converges reliably without destabilizing training.

\subsection{Generalizability: Evaluation on LLVIP Dataset}

To evaluate the broader applicability of our model beyond medical imaging, we tested it on the facades dataset \cite{isola2017image}. The facades dataset is a commonly used benchmark in image-to-image translation tasks. It contains paired images of the façades of architectural buildings and their corresponding label maps. The dataset consists of approximately 506 image pairs (400 for training and 106 for testing). Each image pair typically consists of a label map or semantic segmentation showing structural elements and a target image which is a real photograph of the corresponding building façade.

Table \ref{tab:comparison2} reports the evaluation metrics, demonstrating that our model achieves superior performance compared to Pyramid pix2pix in this domain. The ability of our model to generalize effectively to non-medical image-to-image translation tasks highlights its robustness and adaptability, further validating the impact of our variance-based loss function in mitigating mode collapse across different domains.

These findings emphasize the broader significance of our approach, indicating that our model is not only highly effective for H\&E to IHC translation but also possesses the versatility to be extended to various image translation applications beyond biomedical imaging.


\section{Discussion}
The challenge of accurately translating hematoxylin and eosin (H\&E) stained tissue into immunohistochemistry (IHC) representations has impeded broader clinical deployment of generative models in digital pathology. Existing frameworks often struggle to preserve fine-grained morphological and phenotypic features critical for accurate biomarker assessment. Our work addresses this long-standing limitation by introducing a novel variance-based penalty term into the pyramid pix2pix architecture, which fundamentally reshapes the loss penalty landscape to encourage diversity and fidelity in the generated images. The proposed modification to the loss function, detailed in eq. \ref{eq:ours}, effectively addresses the mode collapse problem by introducing a penalty for generating less diverse images. The results affirm the effectiveness of this modification across three pivotal dimensions: improved HER2-positive translation accuracy, mitigation of mode collapse, and generalizability to non-medical image translation tasks.

\subsection{High-Fidelity Translation of HER2-Positive (IHC 3+) Images}
HER2-positive tumors, marked by strong circumferential membrane staining, present high morphological heterogeneity that has historically confounded generative models. While prior models such as pix2pix and pyramid pix2pix achieved reasonable performance on low-expression IHC classes (0/1+/2+), they consistently failed to generate realistic IHC 3+ images, a shortcoming with direct clinical implications. By incorporating a variance-based penalty that directly targets the model’s tendency toward oversimplified outputs, our method achieves substantial gains in PSNR and SSIM for IHC 3+ cases. Crucially, the 33\% decrease in Fréchet Inception Distance (FID) for IHC 3+ against the state of the art methods, a metric sensitive to perceptual realism and distributional alignment, demonstrates that our model not only improves quantitative performance but also generates images with higher diagnostic plausibility. As shown in Figure \ref{fig:comparison2}, our method consistently produces more diverse and representative images across different classes of input, particularly in comparison to baseline methods such as pix2pix and pyramid pix2pix.

\subsection{Variance-Based Penalty and Mitigation of Mode Collapse}

Mode collapse remains one of the most persistent and debilitating issues in GAN-based translation models, often manifesting itself as repetitive or degenerate outputs. Traditional solutions have explored architectural adjustments or adversarial training stabilization techniques, but few have directly targeted the diversity of outputs through the loss function. Our contribution departs from this paradigm by introducing a batch-level variance regularization term that penalizes the model when the variance in generated images deviates significantly from that of the real distribution. This variance-based constraint acts as a statistical anchor, encouraging the generator to produce outputs that span the intrinsic variability of the target domain.

\begin{figure}[t]
\centering

\begin{minipage}[t]{0.48\textwidth}
\centering
\includegraphics[width=\textwidth]{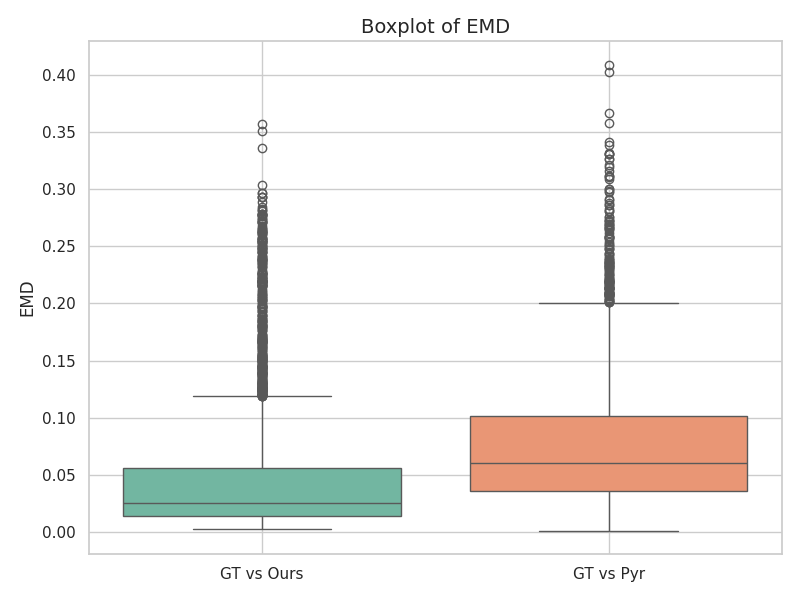}

\vspace{2mm}
{\small \textbf{(a)}}
\label{fig:emd}
\end{minipage}
\hfill
\begin{minipage}[t]{0.48\textwidth}
\centering
\includegraphics[width=\textwidth]{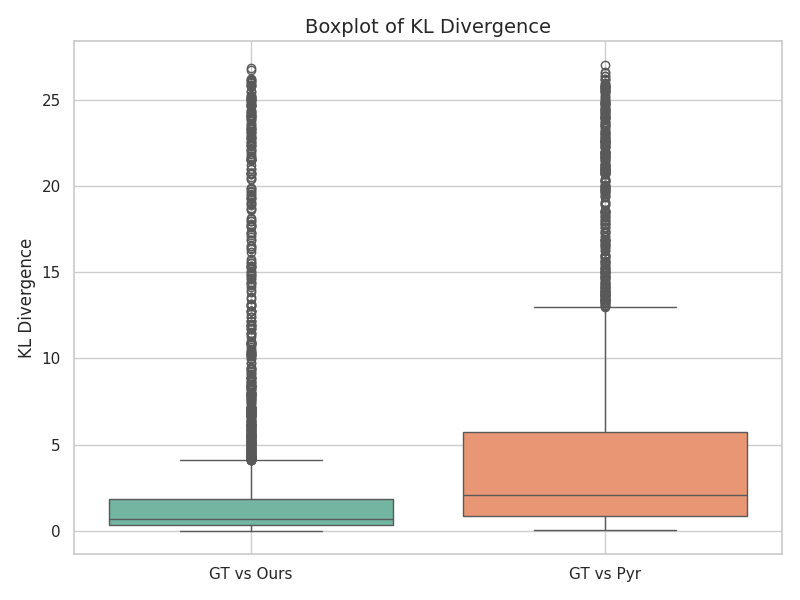}

\vspace{2mm}
{\small \textbf{(b)}}
\label{fig:kl}
\end{minipage}

\caption{Comparison of EMD and KL divergence across models. Boxplot (a) illustrates the distribution of EMD values, and boxplot (b) illustrates the distribution of KL divergence values computed between the color histograms of ground truth (GT) images and those generated by different models: Ours and Pyramid Pix2Pix. The lower median and tighter spread in the GT vs Ours distribution in both boxplots indicate higher similarity in pixel intensities and greater consistency across samples.}
\label{fig:comparison_boxplots}
\end{figure}

 Mode collapse arises when the generator fails to capture certain modes of the real data distribution. Two of the metrics that can quantify the reduction in mode collapse are Earth Mover’s Distance (EMD) \cite{rubner2000earth} and Kullback–Leibler (KL) Divergence \cite{kullback1951information}. EMD quantifies the minimal effort needed to transform one distribution into another. The lower EMD for our model as shown in Figure \ref{fig:comparison_boxplots}(a) indicates that the distribution of generated images closely aligns with that of the ground truth. Similarly, Kullback–Leibler (KL) Divergence measures how one probability distribution diverges from a reference distribution. The lower KL divergence observed in  Figure \ref{fig:comparison_boxplots}(b) for our model suggests a better approximation of the real data distribution and a reduction in mode collapse.

The empirical results confirm the efficacy of this strategy. Across HER2 classes, our model exhibits higher intra-class variance while maintaining or improving fidelity metrics. This balance between diversity and accuracy is critical, especially in histopathological contexts where subtle phenotypic variations bear diagnostic weight. Our findings suggest that loss-level interventions, particularly those informed by global statistical properties, offer a promising new direction for tackling mode collapse in medically constrained GAN applications.

\subsection{Generalizability Beyond Medical Imaging}
The proposed model was initially optimized for histopathological image translation task, we further evaluated its generalization capability on the facades dataset \cite{isola2017image}, a benchmark in the domain of visible-to-infrared translation. The ability of our model to outperform pyramid pix2pix in PSNR, SSIM and FID on this out-of-domain task highlights the robustness and adaptability of our variance-regularized framework. This observation is nontrivial: most domain-specific models overfit to localized feature distributions and exhibit poor transferability. Our results indicate that the variance-based penalty does not merely fine-tune the model for HER2 classification, but rather instills a broader inductive bias toward generative diversity, a quality that enhances performance even in unrelated modalities.

The implications of this work extend beyond technical performance gains. From a clinical standpoint, the improved translation of HER2-positive cases opens avenues for more reliable, cost-effective, and scalable HER2 screening pipelines, especially in low-resource settings where IHC reagents and pathologist expertise are limited. From a methodological perspective, our findings support a more systemic reconsideration of loss functions in GANs, advocating for principled regularization strategies that reflect underlying statistical properties of the data.

Future work may explore hierarchical or region-specific variance constraints to further tailor the loss function to heterogeneous tissue morphologies. Moreover, integrating uncertainty quantification, through Bayesian or ensemble approaches, could provide valuable confidence estimates, enhancing clinical interpretability and trust. Finally, expanding the model to support multiplexed biomarker generation could transform its utility from a diagnostic aid into a tool for digital pathology simulation and augmentation.

\section{Conclusion}

This paper suggests changes in the pyramid pix2pix model for translation of high-fidelty IHC images from H\&E images. IHC images are used for the detection of HER2 cancer. To reduce the problem of mode collapse in the pyramid pix2pix model, we introduce a term in the objective function of the model that will penalize the model for generating images with relatively less variance with respect to the input images. The proposed model not only outperforms the state-of-the-art models in translation from H\&E images to IHC images, it also reports better evaluation scores for generic datasets. In the presence of enough computational resources, the input images do not need to be cropped, which may further improve our results.

\section{Tables}
\begin{table}[htbp]
\centering
\caption{Comparison of Evaluation Metrics for pix2pix, pyramid pix2pix, and Our Method for IHC 0.}

\begin{tabular}{cccc}
\hline
\textbf{Metrics} & \textbf{pix2pix} & \textbf{pyramid pix2pix} & \textbf{ours} \\ \hline
\textbf{PSNR}    & 19.6            & 19.2                  & 20.48        \\
\textbf{SSIM}    & 0.52           & 0.46                  & 0.55         \\
\textbf{FID}     & 914.32          & 920.98               & 890.46      \\ \hline
\end{tabular}%

\label{tab:ihc_0}
\end{table}

\begin{table}[htbp]
\centering
\caption{Comparison of Evaluation Metrics for pix2pix, pyramid pix2pix, and Our Method for IHC 1+.}

\begin{tabular}{cccc}
\hline
\textbf{Metrics} & \textbf{pix2pix} & \textbf{pyramid pix2pix} & \textbf{ours} \\ \hline
\textbf{PSNR}    & 22.22           & 21.07                  & 21.65       \\
\textbf{SSIM}    & 0.47             & 0.44                  & 0.48         \\
\textbf{FID}     & 605.58           & 642.79                 &  472.19     \\ \hline
\end{tabular}%

\label{tab:ihc_1_plus}
\end{table}

\begin{table}[htbp]
\centering
\caption{Comparison of Evaluation Metrics for pix2pix, pyramid pix2pix, and Our Method for IHC 2+.}

\begin{tabular}{cccc}
\hline
\textbf{Metrics} & \textbf{pix2pix} & \textbf{pyramid pix2pix} & \textbf{ours} \\ \hline
\textbf{PSNR}    & 20.9           & 20.4                   & 21.3       \\
\textbf{SSIM}    & 0.44            & 0.42                   & 0.46         \\

\textbf{FID}     & 473.84           & 527.4                & 395.4      \\ \hline
\end{tabular}%

\label{tab:ihc_2_plus}
\end{table}

\begin{table}[htbp]
\centering
\caption{Comparison of Evaluation Metrics for pix2pix, pyramid pix2pix, and Our Method for IHC 3+.}

\begin{tabular}{cccc}
\hline
\textbf{Metrics} & \textbf{pix2pix} & \textbf{pyramid pix2pix} & \textbf{ours} \\ \hline
\textbf{PSNR}    & 19.2          & 19.4                  & 19.5        \\
\textbf{SSIM}    & 0.39            & 0.38                 & 0.42         \\
\textbf{FID}     & 733.01           & 758.74                 & 508.38       \\ \hline
\end{tabular}%

\label{tab:ihc_3_plus}
\end{table}

\begin{table}[htbp]
\centering
\caption{Comparison of Evaluation Metrics for pix2pix, pyramid pix2pix and Our Method using all classes of BCI datatset}
\begin{tabular}{@{}lccc@{}}
\toprule
\textbf{Metric} & \textbf{pix2pix}& \textbf{pyramid pix2pix} & \textbf{ours} \\ \midrule
\textbf{PSNR}  & 20.74 & 21.15 & 22.16 \\
\textbf{SSIM} &0.44  & 0.43 & 0.47 \\
\textbf{FID}  & 472.6 & 516.75 & 346.37 \\ \bottomrule
\end{tabular}
\label{tab:avg_comparison}
\end{table}




\begin{table}[htbp]
\centering
\caption{Comparison of evaluation metrics for Pyramid Pix2Pix and the proposed method on the Facades dataset.}
\label{tab:comparison2}

\begin{tabular}{lcc}
\toprule
\textbf{Metric} & \textbf{Pyramid Pix2Pix} & \textbf{Ours} \\
\midrule
PSNR & 12.91 & 12.80 \\
SSIM & 0.24  & 0.25  \\
FID  & 611.34 & 549.07 \\
\bottomrule
\end{tabular}

\end{table}

\section*{Conflict of Interest Statement}

The authors declare that the research was conducted in the absence of any commercial or financial relationships that could be construed as a potential conflict of interest.
\section*{Author Contributions}

SR (Sara Rehmat): Conceptualization, methodology, software development, data curation, formal analysis, validation, visualization, and writing, original draft preparation.

HR (Hafeez-ur-Rehman): Conceptualization, supervision, methodology refinement, investigation, resources, writing—review and editing, and project administration.

BGK (Byeong-Gwon Kang): Conceptualization, supervision, methodology guidance, validation, writing and editing, and funding acquisition.

SA (Sarra Ayouni): Data curation, investigation, validation, and writing, review and editing.

YN (Yunyoung Nam): Supervision, resources, writing, review and editing, validation and funding acquisition.

\section*{Funding}
This work was supported by the National Research Foundation of Korea(NRF) grant funded by the Korea government(MSIT) (No. RS-2023-00218176) and the Soonchunhyang University Research Fund. Authors like to thanks Princess Nourah bint Abdulrahman University Researchers Supporting Project number (PNURSP2026R896), Princess Nourah bint Abdulrahman University, Riyadh,Saudi Arabia.

\section*{Acknowledgments}
The authors acknowledge the Department of Computer Science, National University of Computer and Emerging Sciences (FAST-NUCES), Peshawar Campus, Pakistan, for providing hardware support and the necessary laboratory resources. We also thank the BCI dataset providers for making the histopathology images publicly available, which enabled this research. Computational resources were supported by Google Colaboratory (Google Colab).

\section*{Data Availability Statement}
The datasets used and/or analyzed during the current study are available from the corresponding author upon request.


   



\bibliographystyle{Frontiers-Vancouver} 
\bibliography{references}

\end{document}